\documentclass[fleqn,usenatbib]{mnras}
\usepackage{comment}
\usepackage{gensymb}
\usepackage{url}
\usepackage{newtxtext,newtxmath}
\usepackage[T1]{fontenc}
\usepackage{ae,aecompl}

\usepackage[utf8]{inputenc}
\usepackage{graphicx}	% Including figure files
\usepackage{amsmath}	% Advanced maths commands
\usepackage{subcaption} % Sub-plot figures
\usepackage{lineno}
\usepackage{xcolor}
\usepackage{ulem}

\usepackage{threeparttable}

\usepackage{rotating} % Add this to your preamble
\usepackage{siunitx}
\usepackage{booktabs}
\usepackage{rotating}

\usepackage{adjustbox}
\usepackage{pdflscape}
\usepackage{float}  % Add this in the preamble
\usepackage{newunicodechar}
\newunicodechar{−}{\ensuremath{-}}
\newcommand{\hii}{$\textrm{H}\scriptstyle\mathrm{II}$}
\newcommand{\rev}[1]{#1}

\title[$\ion{H}{ii}$ regions in the SMGPS]{SMGPS: A study of Galactic {$\textrm{H}\scriptstyle\mathrm{II}$} regions with extended morphology}

\author[Ramaila et al.]{Athanaseus J.~T. Ramaila$^{1,2}$\thanks{E-mail:aramaila@sarao.ac.za},
     Mark A. Thompson$^{3,8}$,
     Oleg M. Smirnov$^{1,2,4,11,12}$, \and
     Sphesihle Makhathini$^{5,10}$,
     James O. Chibueze$^{6,7}$, Willice Obonyo$^{6}$,\and
     Chukwuebuka J. Ugwu$^{6}$,
     Cristobal Bordiu$^{8}$, Simone Riggi$^{8}$, Alessio Traficante$^{9}$\\
$^{1}$ Department of Physics and Electronics, Rhodes University, PO Box 94, Makhanda 6140, South Africa \\
$^{2}$ South African Radio Astronomy Observatory, 2 Fir Street, Black River Park, Observatory, Cape Town 7925, South Africa\\
$^{3}$ School of Physics and Astronomy, University of Leeds, Woodhouse Ln, Leeds LS2 9JT, United Kingdom
\\
$^{4}$Institute for Radioastronomy, National Institute of Astrophysics (INAF IRA), Via Gobetti 101, 40129 Bologna, Italy \\
$^{5}$ School of Physics, University of the Witwatersrand, Johannesburg, Braamfontein, 2000, South Africa \\
$^{6}$UNISA Centre for Astrophysics and Space Sciences (UCASS),
College of Science, Engineering and Technology, University of South Africa, Cnr Christian\\ de Wet Rd and Pioneer Avenue, Florida Park, 1709, Roodepoort, South Africa.\\
$^{7}$ Department of Physics and Astronomy, Faculty of Physical Sciences, University of Nigeria, Carver Building, 1 University Road, Nsukka 410001, Nigeria \\
$^{8}$ INAF - Osservatorio Astrofisico di Catania, Via Santa Sofía 78, I−95123 Catania, Italy \\
$^{9}$ Istituto di Astrofisica e Planetologia Spaziali (IAPS), INAF, Via Fosso del Cavaliere 100, 00133 Roma, Italy \\
$^{10}$ INAF - Osservatorio Astronomico di Cagliari, Via della Scienza 5, 09047 Selargius (CA), Italy \\
$^{11}$Astrophysics, Department of Physics, University of Oxford, Keble Road, Oxford, OX1 3RH, UK\\
$^{12}$Breakthrough Listen, Astrophysics, Department of Physics, The University of Oxford, Keble Road, Oxford, OX1 3RH, UK\\
}

\date{Accepted XXX. Received YYY; in original form ZZZ}

\pubyear{2026}

\begin{document}
%\linenumbers
\label{firstpage}
\pagerange{\pageref{firstpage}--\pageref{lastpage}}
\maketitle

\begin{abstract}
We present a study of ionised hydrogen ({$\textrm{H}\scriptstyle\mathrm{II}$}) regions in the Galactic Plane using data from the SARAO MeerKAT Galactic Plane Survey (SMGPS). The SMPGS is a wide-field, wide-band 1.3 GHz radio continuum survey ($251^\circ \leq l \leq 358^\circ$ and $2^\circ \leq l \leq 61^\circ$ at $\quad |b| \leq 1^\circ.5$) that has enabled us to trace the diffuse emission enveloping recently formed massive stars. Our multifrequency synthesis images reveal faint and extended emission that was previously overlooked by {$\textrm{H}\scriptstyle\mathrm{II}$} region surveys. We report the distances and Lyman-photon flux ($N_{\mathrm{Ly}}$) measurements for 1,327 Galactic {$\textrm{H}\scriptstyle\mathrm{II}$} regions from which we characterise the spectral types for candidate ionising stars. The spectral types range from \rev{B2} to O4. The typical stellar spectral type responsible for ionisation is the B0, which constitutes about $\text{16\%}$ of our catalogue, corresponding to a mean $\log(N_{\mathrm{Ly}}) = 47.5\ \mathrm{s}^{-1}$. Moreover, as a result of the lack of radio recombination line (RRL) velocity measurements for faint {$\textrm{H}\scriptstyle\mathrm{II}$} regions, we identify the effective completeness limit at $\log(N_{\mathrm{Ly}}) \approx 47.6\  \mathrm{s}^{-1}$. The multiwavelength approach reveals that the physical radius at 1.3 GHz and in the mid-infrared are well correlated \rev{with a slope of $1.15 \pm 0.02$}. We find clear power-law relations between $N_{\mathrm{Ly}}$ and physical radius, and an inverse correlation between electron density and radius ($n_{\rm e} \propto R^{-0.73}$). However, no significant correlation is observed between the $N_{\mathrm{Ly}}$ and Galactocentric distance, suggesting that the observed trends are governed primarily by local star-forming environments rather than large-scale Galactic gradients.

\end{abstract}


%%%%%%%%%%%%%%%%%%%%%%%%%%%%%%%%%%%%%%%%%%%%%%%%%%

%%%%%%%%%%%%%%%%% BODY OF PAPER %%%%%%%%%%%%%%%%%%

\section{Introduction}

\rev{Star formation is a fundamental astrophysical process that governs the evolution of galaxies and significantly influences the large-scale structures and dynamics of the universe.} When large clouds of interstellar gas collapse, they lead to the birth of stars and star clusters. The more massive stars ($\rev{>}8-10$ solar masses), which have high surface temperatures (ranging from $\sim30,000$ to $50,000$ K), emit intense ultraviolet (UV) radiation. 
The UV photon energies exceed the \SI{13.6}{\electronvolt} \rev{required to ionise atomic hydrogen from its ground state}. This process results in the \rev{photoionisation} of the surrounding hydrogen in the interstellar medium (ISM), giving rise to \hii~regions \citep{spitzer1978}.
\rev{These massive stars, typically found in OB associations, belong to spectral types O and B, and} are the primary tracers of spiral patterns and the overall distribution of hydrogen gas in galactic discs. 
\rev{Broadband radio-frequency} observations provide \rev{a powerful tool for} examining the large-scale distribution of \rev{the Galactic ISM}. \rev{While millimetre-wave spectroscopy traces the cold} molecular clouds \citep[e.g.,][]{1970ApJ...161L..43W,2015ARA&A..53..583H}, \rev{centimetre-wave continuum observations (such as those presented here) effectively trace the ionised} plasma. \rev{Crucially, both regimes} bypass the extreme dust obscuration inherent to the Galactic Plane \rev{that hinders optical surveys}.
At frequencies around 1\ GHz, radio continuum emission primarily arises from two \rev{distinct} mechanisms: thermal \rev{bremsstrahlung (free-free)} emission produced by electrons in \rev{the} plasma surrounding hot, massive stars (such as $\text{O}$ and $\text{B}$ types, and their evolved counterparts like Wolf-Rayet stars and Luminous Blue Variables) and non-thermal \rev{synchrotron emission (also a form of bremsstrahlung)} generated by relativistic electrons spiralling \rev{along} magnetic fields. Thermal free-free emission, typical of classical {$\textrm{H}\scriptstyle\mathrm{II}$} regions, exhibits an almost flat spectrum with $\alpha \approx -0.1$ when optically thin and a positive spectral index of up to $\sim 2$ in the optically thick regime \citep{rybicki_lightman_1979}. \hii~regions are distributed throughout the Galactic disc and are among the brightest Galactic objects at radio and mid-infrared wavelengths. Unlike other indicators of \rev{G}alactic star formation, the clear identification of an \hii~region directly points to the presence of massive star formation \citep{1997pism.book.....D,2002ARA&A..40...27C,2014ApJS..212....1A}. This study aims to characterise the population of \hii~regions in the Galaxy by analysing their electron densities and physical sizes, determined using the derived kinematic distances. We also explore the distribution of ionising photon fluxes and candidate ionising stars throughout the Galactic disc and evaluate the completeness of the known \hii~regions sample. A comprehensive survey of the Galactic \hii~regions is critical to testing and refining models of the \rev{structure and kinematics of the Milky Way \citep[e.g.][]{2019ApJ...885..131R}, as these regions serve as primary tracers of the spiral arm architecture.}

A range of electron densities is observed in \hii~regions, with the most compact regions exhibiting the highest values. As an \hii~region expands and interacts with the surrounding medium, its electron density typically decreases. \hii~regions are classified according to size, electron density and evolutionary stage \citep{1990A&ARv...2...79C}. The youngest and most compact phase corresponds to hypercompact {$\textrm{H}\scriptstyle\mathrm{II}$} regions (HC{$\textrm{H}\scriptstyle\mathrm{II}$}), characterised by diameters smaller than 0.1 pc and electron densities exceeding  $10^5 \rm \, cm^{-3}$ \rev{\citep{2023MNRAS.524.4384P,2024MNRAS.533.2005P,2025MNRAS.538.2267P,2019MNRAS.482.2681Y,2021A&A...645A.110Y}}. These extremely dense ionised regions form around the youngest massive protostars, remaining deeply embedded within their natal molecular cores. At the other extreme, classical \hii~regions represent the most evolved stage, spanning tens to hundreds of parsecs with electron densities typically below $10^3 \rm \, cm^{-3}$ \rev{(e.g. \citealt{2026A&A...706A.280K})}. The more extensive regions host mature stellar populations whose intense radiation and stellar winds significantly influence the surrounding interstellar medium. \rev{This stellar feedback can regulate future star formation or trigger the collapse of nearby molecular clumps \citep{2012MNRAS.421..408T,2013MNRAS.435.2003H,2021A&A...656A.101M}.} The lifecycle unfolds from dense cradles of {$\textrm{HCH}\scriptstyle\mathrm{II}$}, ultra-compact ({$\textrm{UCH}\scriptstyle\mathrm{II}$}), compact ({$\textrm{CH}\scriptstyle\mathrm{II}$}), and ultimately classical {$\textrm{H}\scriptstyle\mathrm{II}$} regions, with a corresponding decrease in density as they expand into their environments. {$\textrm{UCH}\scriptstyle\mathrm{II}$} are extremely dense regions, with typical physical sizes of $0.05 < R \lesssim 0.5$ pc and often optically thick at 1 GHz. {$\textrm{CH}\scriptstyle\mathrm{II}$} are moderately dense, roughly $0.5 < R \lesssim 2$ pc \citep{2002ARA&A..40...27C}.

Focused searches for Galactic {$\textrm{H}\scriptstyle\mathrm{II}$} regions began around 70 years ago with photographic plate surveys in the northern hemisphere by \citet{1953ApJ...118..362S} and the southern hemisphere by \citet{1955MmRAS..67..155G}. The anticipation \citep{1959SvA.....3..813K} and detection \citep{1965Sci...150..339H} of radio recombination lines (RRLs) presented a fresh, unobstructed spectroscopic marker for Galactic {$\textrm{H}\scriptstyle\mathrm{II}~$} regions. In subsequent decades, RRL surveys have revealed hundreds of novel nebulae, most of which are optically obscured \citep{1970A&A.....4..357R, 1970ApL.....5...99W, 1970A&A.....4..357R, 1980A&AS...40..379D, 1987A&A...171..261C}. For a comprehensive review of the history of RRL surveys in {$\textrm{H}\scriptstyle\mathrm{II}$} regions, refer to \citet{2019ApJS..240...24W}. The use of RRLs has been vital for studying {$\textrm{H}\scriptstyle\mathrm{II}$} regions, as they provide essential line-of-sight velocities for calculating kinematic distances and allow for the direct determination of the ionised gas electron temperature ($T_{\rm e}$). Modern large-scale efforts \rev{have significantly expanded the census of Galactic star formation. The GBT \hii~Region Discovery Survey (HRDS; \citealt{2011ApJS..194...32A,2018ApJS..234...33A}) effectively doubled the number of known \hii~regions in the northern Galactic plane with 887 discoveries, while its southern counterpart, the SHRDS (\citealt{2019ApJS..240...24W}), provided 275 new RRL detections, including 148 previously unconfirmed nebulae. These surveys have been instrumental in} providing crucial velocity constraints and associated $T_{\rm e}$ estimates \rev{for nearly a thousand new sources}.

In addition to radio continuum and recombination line observations, \hii~regions are also identified in the mid-infrared, such as the WISE \hii~region catalogue by \citet{2014ApJS..212....1A}. The comparison of these datasets enhances our understanding of the dust and molecular gas structure surrounding the ionising stars through a comprehensive multiwavelength study \rev{\citep{2013MNRAS.435..400U,2014MNRAS.443.1555U,2018MNRAS.473.1059U}}. An all-sky mid-infrared survey was carried out by NASA's Wide-field Infrared Survey Explorer (WISE) during its primary mission in 2010–2011 \citep{2010AJ....140.1868W, 2011ApJ...731...53M}. The WISE images provide mid-infrared coverage at 3.4, 4.6, 12, and 22\ $\mu$m, offering valuable insight into thermal dust emission and the presence of Polycyclic aromatic hydrocarbon (PAH) signatures that trace photodissociation regions around the ionised gas. The WISE \hii~region catalogue (V2.3) contains 8,412 objects, comprising 2,210 known \hii~regions with detected RRL emission and 6,202 candidate or radio-quiet regions identified from their distinctive infrared morphologies \rev{\citep{2014ApJS..212....1A}}. This dataset complements our radio observations by tracing the dust and molecular environments associated with the ionised gas.
%\footnote{\url{https://astro.phys.wvu.edu/wise/}}

Recent radio observations of \hii~regions show remarkable congruence with their infrared counterparts regarding the spatial extent and infrared-to-radio flux density ratios \citep{2017ApJ...846...64M}. This alignment is primarily attributed to advancements in observational sensitivity and resolution of modern radio telescopes. Additionally, high-resolution radio and mid-infrared observations have revealed that some \hii~regions possess compact cores embedded within extended diffuse ionised halos that can span up to several parsecs \citep{2001ApJ...549..979K,2024A&A...689A.254D}.
Typically, infrared emissions are observed to extend beyond the radio-continuum boundaries of compact {$\textrm{H}\scriptstyle\mathrm{II}$} regions. 
This suggests that broader infrared-emitting dust encompasses the ionised radio structures. These findings underscore the agreement between the detected radio emissions and the surrounding dust revealed at mid-infrared wavelengths.

Despite decades of multiwavelength investigations, various unresolved questions persist, for instance, the details of stellar evolution, especially concerning the later stages of massive stars \citep{2011BSRSL..80..335U}. Using the MeerKAT telescope, which commenced its early science phase in 2016 \citep{2016mks..confE...1J}, the SARAO MeerKAT 1.3 GHz Galactic Plane Survey (SMGPS) began in 2018 to comprehensively examine the entire southern Galactic plane at 1283\ MHz, achieving a deep continuum survey with a typical rms of $\sim$ 15 $\mu$Jy/beam and an angular
resolution of 8 arcsec \citep{2024MNRAS.531..649G}. It has generated highly sensitive wide-field maps of Galactic radio continuum emission. However, since our sample is drawn from the SMGPS extended source catalogue, with sources angular sizes at least five beams, the effective minimum angular radius is $\sim20$ arcsec. This corresponds to a physical scale of $\sim 0.5$ pc at a distance of 5 kpc, and $\sim 1$ pc at 10 kpc. Consequently, our data are ideally suited for a comprehensive census of classical $\textrm{H}\scriptstyle\mathrm{II}$ regions, providing sufficient resolution to resolve their internal structure and diffuse emission across the Galactic plane. \rev{While recent high-resolution surveys such as CORNISH \citep{Hoare2012_CORNISH_Design, Purcell2013_CORNISH_Catalog} and GLOSTAR \citep{2021A&A...651A..85B} have provided unprecedented views of compact and ultra-compact \hii~regions, they are often less sensitive to large-scale, low-surface-brightness emission. On smaller scales, the Search for Clandestine Optically Thick Compact \hii\ Regions survey \citep[SCOTCH;][]{2023MNRAS.524.4384P} concentrates on the poorly understood hypercompact \hii\ region phase. Our study, using 1.3 GHz SMGPS data, complements these by focusing on the extended phase of \hii~region evolution, using the unique low-surface-brightness capabilities of MeerKAT.}

Our analysis highlights a distinctive characteristic of these resolved {$\textrm{H}\scriptstyle\mathrm{II}$} regions: their association with a physically extensive ionised gas that extends beyond the initially identified core structures and their distribution within the \rev{G}alactic disc. \rev{The paper is structured as follows:} Section \ref{obs} outlines the SMGPS observations and the data products that were used. In Section \ref{catalogue}, the sample selection process is described, beginning with the SMGPS extended source catalogue \citep{2025A&A...695A.144B}. Section \ref{results} provides the results of the Galactic {$\textrm{H}\scriptstyle\mathrm{II}$} region sample with extended morphology observed with the MeerKAT telescope.
\rev{I}n Section \ref{discussion}, we discuss the derived properties of these $\textrm{H}\scriptstyle\mathrm{II}$ regions, demonstrating how their ionising output scales with physical size across the classical regime and examining the factors—such as the local environment versus Galactic location—that govern their dynamical evolution.
\rev{Finally, Section \ref{conclusion} summarises our main findings and discuss the implications of this study for future high-sensitivity Galactic plane surveys.}

\section{SARAO MeerKAT Galactic Plane Survey}
\label{obs}
The SARAO MeerKAT 1.3 GHz Galactic Plane Survey (SMGPS) is currently the deepest radio continuum survey \rev{of the Galactic Plane} at a frequency of $\sim$ 1 GHz. The observing band was divided into 4096 spectral channels, with an 8-second integration time, and spanned the frequency range from 856 to 1712\ MHz.
These observations span a two-year period, from July 21st, 2018, to March 14th, 2020, covering almost half of the Galactic Plane, encompassing the two main longitude ranges $251^\circ \leq l \leq 358^\circ$ and $2^\circ \leq l \leq 61^\circ$ at $\quad |b| \leq 1^\circ.5$ \citep{2024MNRAS.531..649G}. Each mosaic achieved a synthesised beam of 8 arcsec and a background sensitivity of about 15\ $\mu$Jy/beam with slightly higher noise towards regions with bright sources. The survey delivers astrometric precision of approximately 0.5 arcsec, with flux densities accurate to within 5\%, benchmarked against other surveys.
\newline
The SMGPS observations and data are fully described in \rev{\citet{2024MNRAS.531..649G}}. Subsequent papers detail catalogues of extended, filamentary, and compact/point sources, respectively \citep{2025A&A...695A.144B,2025MNRAS.536.1428W,2026MNRAS.546f1849M}. In this work, we use the mosaiced SMGPS data cubes and the extended source catalogue derived from the data cubes \citep{2025A&A...695A.144B} to study the radio emission from classical {$\textrm{H}\scriptstyle\mathrm{II}$} regions. \rev{The precise selection criteria and subsequent cross-matching framework are outlined in the following subsections.}

\begin{table}
\centering
\caption{Summary of SMGPS {$\textrm{H}\scriptstyle\mathrm{II}$} Region Sample Reduction.}
\begin{tabular}{lc}
\toprule
\textbf{Selection Step} & \textbf{Count} \\
\midrule
SMGPS Extended Source Catalogue & 16,534 \\
Initial $ \text{SMGPS}-\text{WISE}$ {$\textrm{H}\scriptstyle\mathrm{II}$} Associations & 3,323 \\
Morphological Criteria Excluding Groups  & 3,039 \\ 
Single LSR Velocity Detections & 1,327 \\
\midrule
\label{tab:summay}
\end{tabular}
\end{table}

\begin{figure}
    \centering
    \includegraphics[width=\columnwidth]{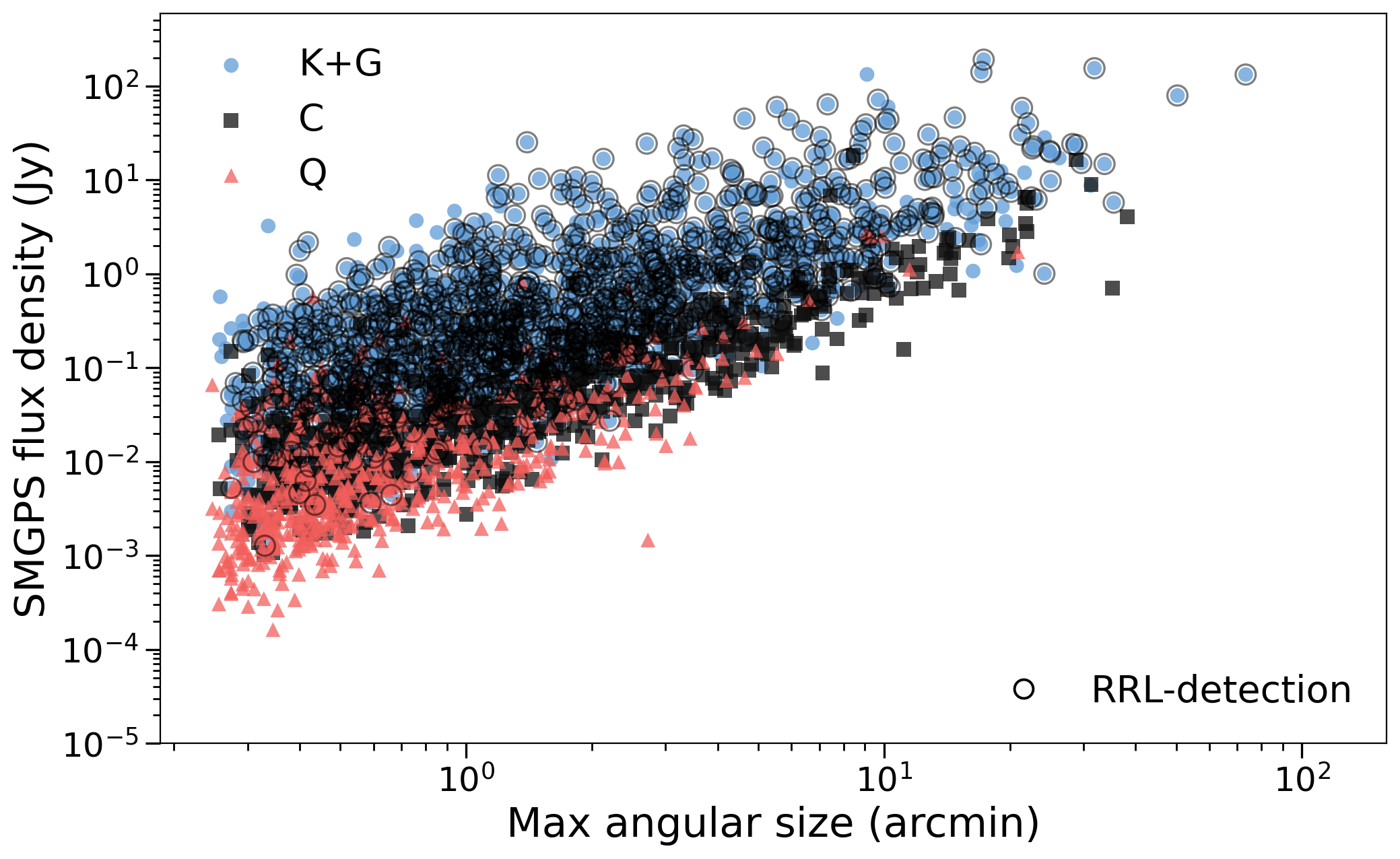}
    \caption{\rev{Distribution of the SMGPS \hii\ region sample in the flux density–angular size plane, adapted from \citet{2025A&A...695A.144B}. The background symbols indicate the parent classifications: high-confidence known and candidate regions (K+G; circles), compact candidates (C; squares), and quiet candidates (Q; triangles). The open circles highlight the subset of sources with confident Radio Recombination Line (RRL) detections analysed in this work. This overlay demonstrates that our RRL-detected sample broadly samples the parameter space of the parent K+G catalogue across more than four orders of magnitude in radio flux.}}
    \label{fig:hii_population}
\end{figure}

\section{SMGPS Sample Selection}
\label{catalogue}

The source extraction process followed the methods described by \citet{2016MNRAS.460.1486R, 2019PASA...36...37R} using the {\tt{caesar}} source finder. Subsequent identification and characterisation of the extended emission involved multi-Gaussian fitting and the application of selection criteria based on morphological properties to produce a catalogue of extended sources \citep{2025A&A...695A.144B}. Each source is assigned a distinct identification number and classified as extended or diffuse according to the following criteria.

Extended sources are large-scale formations with clearly defined limits or regions that extend beyond five synthesised beams. In contrast, diffuse sources are large-scale structures with ambiguous borders and irregular morphologies. Lastly, sources that consist of several different emission components yet have the same underlying astrophysical identity are referred to as multi-island sources, such as radio galaxies and group \hii~regions. Our study is part of a larger effort to systematically catalogue complex emissions within SMGPS; therefore, we focus specifically on the analysis of the extended source catalogue \citep{2025A&A...695A.144B}. The data products include a catalogue\footnote{\url{https://doi.org/10.48479/t1ya-na33}} consisting of 16,534 extended and diffuse sources, covering approximately 500 deg$^2$ in the first, third, and fourth Galactic quadrants. By cross-matching the extracted sources with the \citet{2014ApJS..212....1A} WISE {$\textrm{H}\scriptstyle\mathrm{II}$} region catalogue, we identified 3,323 associations of  extended SMGPS sources with {$\textrm{H}\scriptstyle\mathrm{II}$} regions.

Starting from the initial set of 3,323 associations \citep{2025A&A...695A.144B}, which include known (K), group (G), candidate, radio-quiet WISE {$\textrm{H}\scriptstyle\mathrm{II}$} region identifications. We applied selection criteria to construct a robust sample suitable for determining ionising photon fluxes with the MeerKAT observations, resulting in a catalogue of 1,327\ known {$\textrm{H}\scriptstyle\mathrm{II}$} regions derived from the SMGPS. The primary criteria applied were as follows:

\begin{itemize}
\item \textbf{Morphological Correspondence:} We selected sources that exhibit a clear one-to-one positional correspondence between their radio ($\SI{1.3}{\giga\hertz}$ SMGPS) and mid-infrared ($\SI{12}{\micro\metre}$ WISE) counterparts by excluding group sources so as to limit our analysis to unconfused regions. This leaves only 3,037 known, radio-quiet, and candidate WISE {$\textrm{H}\scriptstyle\mathrm{II}$} regions with SMGPS 1.3 GHz continuum flux measurements.
\newline
\item \textbf{Velocity Reliability:} Accurate distance estimation relies on a reliable $\text{RRL}$ velocity measurement. From this subset, we selected all {$\textrm{H}\scriptstyle\mathrm{II}$} regions with a single local standard of rest velocity ($\text{V}_{\text{LSR}}$) measurement in the \citet{2014ApJS..212....1A}  {WISE $\textrm{H}\scriptstyle\mathrm{II}$} region catalogue\footnote{\url{https://astro.phys.wvu.edu/wise/}}, totaling 1,327 sources. This restricts the final sample to \rev{`}known' {$\textrm{H}\scriptstyle\mathrm{II}$} regions by definition. The requirement of a single $\text{V}_{\text{LSR}}$ avoids ambiguity in determining the distance and ensures a consistent placement of sources in the Galactic context.

\end{itemize}

\begin{figure}
    \centering
    \includegraphics[width=\columnwidth]{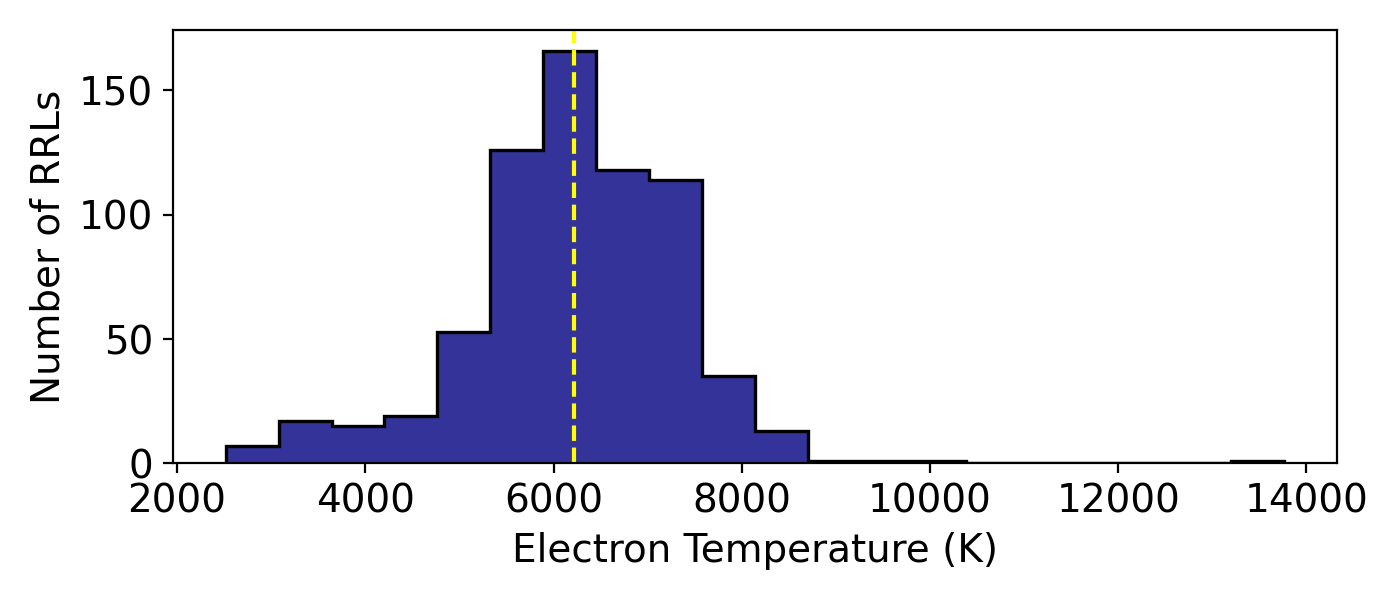}
    \caption{Histogram showing the distribution of electron temperatures ($T_{\rm e}$) available from RRL observations of \hii~regions. The dashed line indicates the mean electron temperature of the observed sample, found to be 6,200 K.}
    \label{fig:te}
\end{figure}

\rev{Furthermore, to evaluate the representativeness of our final RRL-detected sample, we compared its distribution in the flux–angular size plane to the broader parent catalogues. Our RRL-detected sample is by definition restricted to the high-confidence HII regions ($K+G$ sources) from \citet{2025A&A...695A.144B}. These $K+G$ sources occupy a distinct region of higher radio flux density compared to the $Q$ (radio-quite) sources. Notably, our RRL detections span the full range of radio flux densities ($\sim$1.3 mJy to $\sim$190 Jy) and angular scales (0.3' to $\sim$73') present in the $K+G$ population (see Figure~\ref{fig:hii_population}). This distribution demonstrates that our RRL-detected sample probes the diversity of the known \hii~regions within the SMGPS footprint, rather than being limited to only the most compact or brightest subsets.} \rev{Additionally, we} performed a cross-match with an $\text{arcmin}$ tolerance against the \citet{2019ApJS..240...24W} SHRDS catalogue\footnote{\url{https://doi.org/10.7910/DVN/NQVFLE}}, an ATCA radio recombination line and continuum survey of Galactic \hii~regions, to extract electron temperatures ($T_{\rm e}$). 
\rev{The HRDS and SHRDS catalogues provide the kinematic and thermal context for our sample. The HRDS was conducted using the Green Bank Telescope (GBT) with a beam size of $\sim 82''$ at 9~GHz, primarily targeting ionised gas through single-dish pointings with a sensitivity of $\sim 1$--$2$~mJy. In contrast, the SHRDS utilised the ATCA interferometer in the 4--10~GHz range, providing higher angular resolution (up to $\sim 10''$) through multi-frequency synthesised maps, allowing for a more specific localisation of RRL emission.}
Since only 687 sources have measured electron temperatures, we assumed a median temperature value of 6,200 $\text{K}$ \rev{(with interquartile range of $5,605\text{--}6,982\text{ K}$)}, which falls within the typical range for ionised hydrogen gas. The electron temperature distribution is shown in Figure~\ref{fig:te}.

The results of the sample selection process are summarised in Table \ref{tab:summay}. Of the 3,323 initial associations, 1,996 were excluded due to a complete lack of available $\text{RRL}$ velocity information. The characterisation of these objects will require RRL spectroscopic observations to determine reliable kinematic distances. Moreover, almost half lack $T_{\rm e}$ measurements, and therefore, we rely on the mean value of 6,200 $\text{K}$. In the final catalogue of 1,327 sources, we characterise the SMGPS {$\textrm{H}\scriptstyle\mathrm{II}$} region distances, physical radii, Lyman-photon flux, electron density, and the ionising stellar candidate type (Table \ref{tab:hii_sample}).

\begin{figure*}
  \centering
  \begin{subfigure}{0.3\textwidth}
    \includegraphics[width=\linewidth]{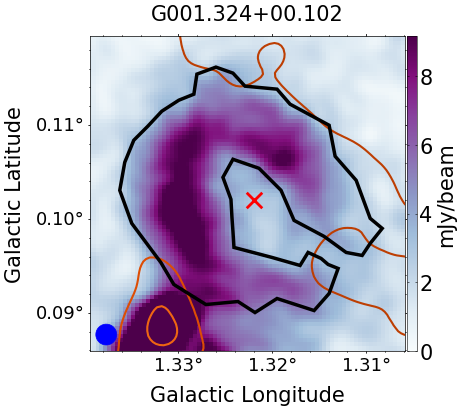}
    \label{fig:hii1}
  \end{subfigure}%
  \hspace{1em}%
  \begin{subfigure}{0.3\textwidth}
    \includegraphics[width=\linewidth]{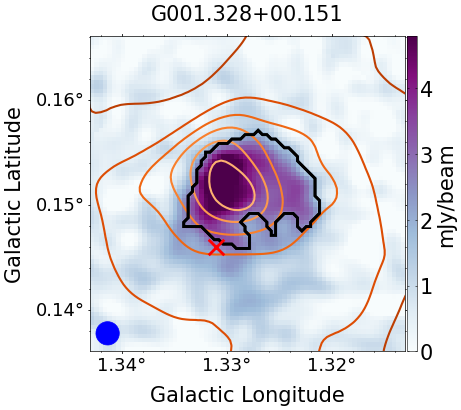}
    \label{fig:hii2}
  \end{subfigure}%
  \hspace{1em}%
  \begin{subfigure}{0.3\textwidth}
    \includegraphics[width=\linewidth]{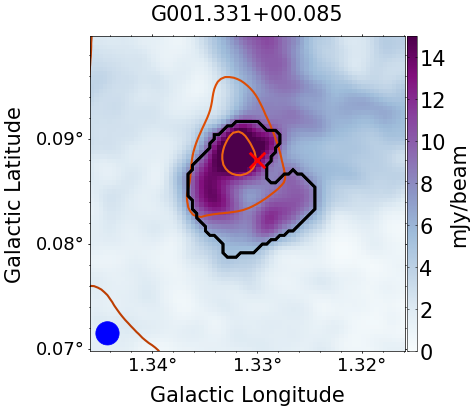}
    \label{fig:hii3}
  \end{subfigure}%
  \hspace{1em}

  \begin{subfigure}{0.3\textwidth}
    \includegraphics[width=\linewidth]{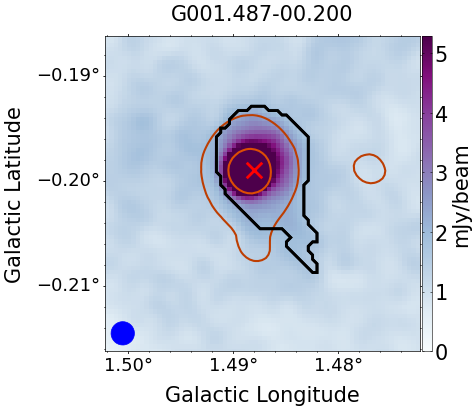}
    \label{fig:hii4}
  \end{subfigure}%
  \hspace{1em}%
  \begin{subfigure}{0.3\textwidth}
    \includegraphics[width=\linewidth]{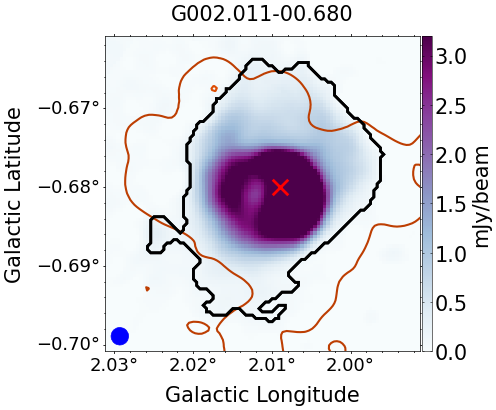}
    \label{fig:hii5}
  \end{subfigure}%
  \hspace{1em}
  \begin{subfigure}{0.3\textwidth}
    \includegraphics[width=\linewidth]{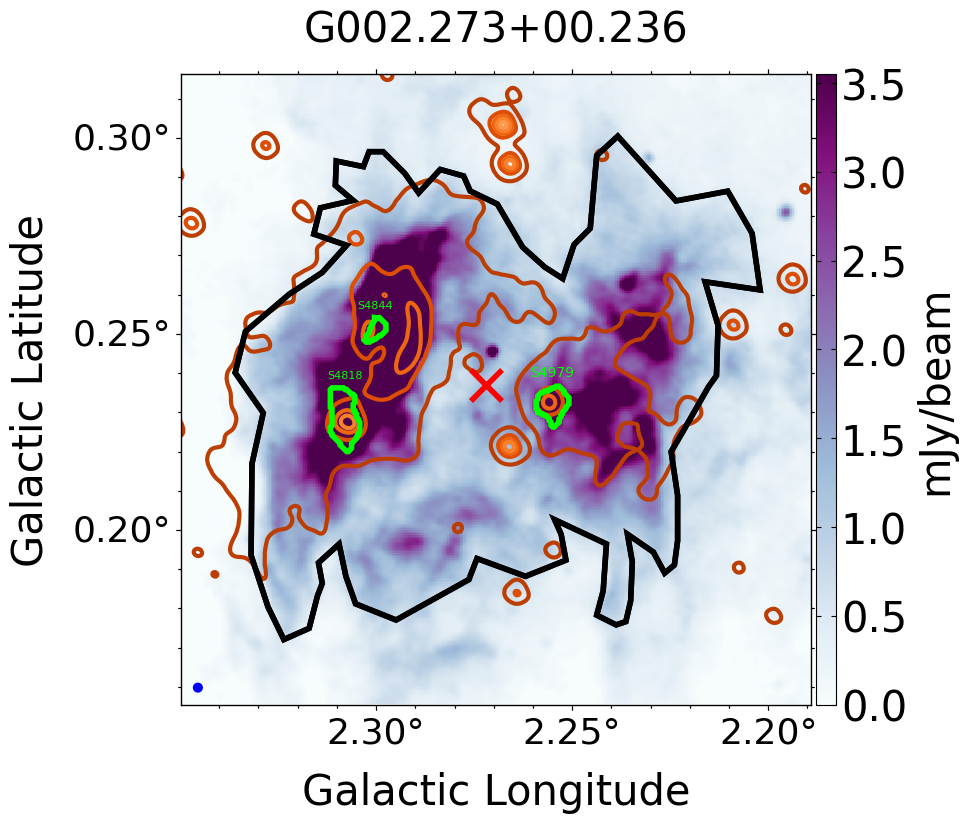}
    \label{fig:hii6}
  \end{subfigure}%
  \hspace{1em}%

  \begin{subfigure}{0.3\textwidth}
    \includegraphics[width=\linewidth]{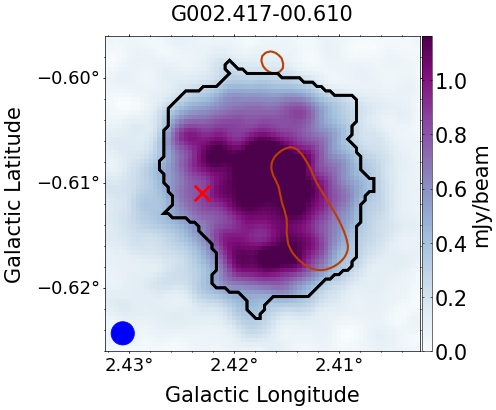}
    \label{fig:hii7}
  \end{subfigure}%
  \hspace{1em}%
  \begin{subfigure}{0.3\textwidth}
    \includegraphics[width=\linewidth]{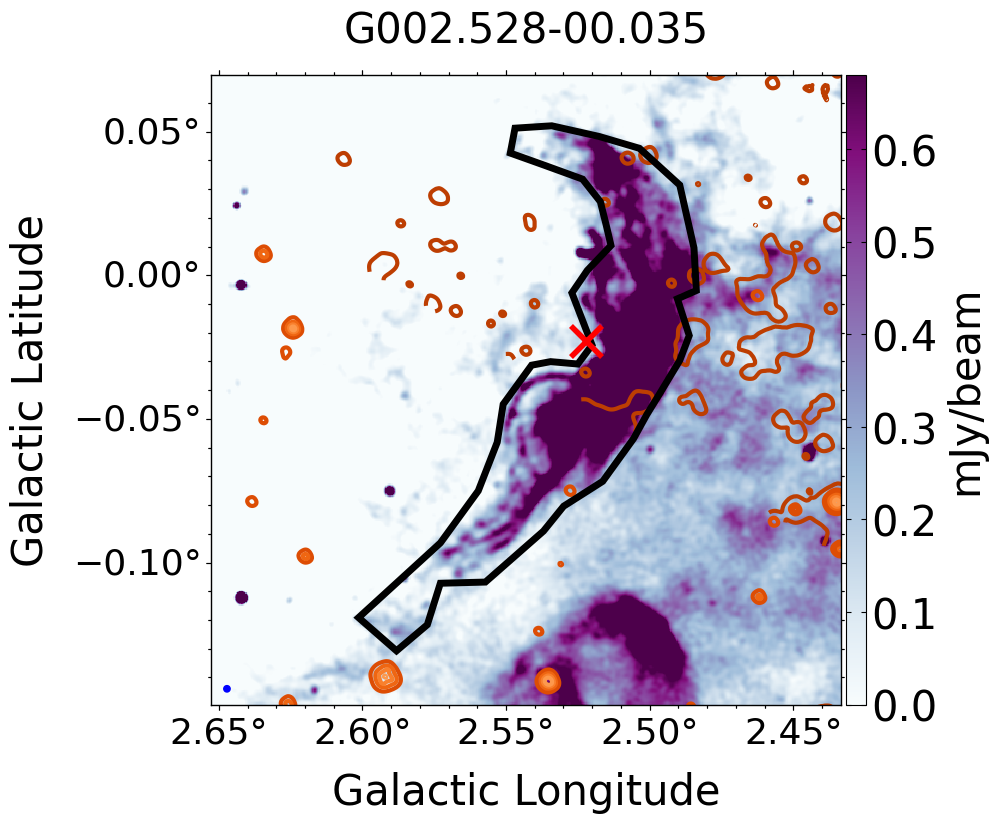}
    \label{fig:hii8}
  \end{subfigure}%
  \hspace{1em}%
  \begin{subfigure}{0.3\textwidth}
    \includegraphics[width=\linewidth]{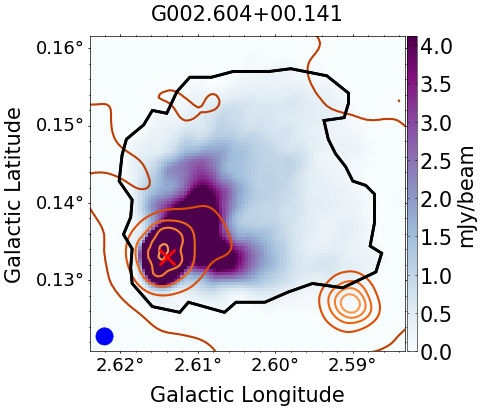}
    \label{fig:9}
  \end{subfigure}%
  \hspace{1em}%

  \begin{subfigure}{0.3\textwidth}
    \includegraphics[width=\linewidth]{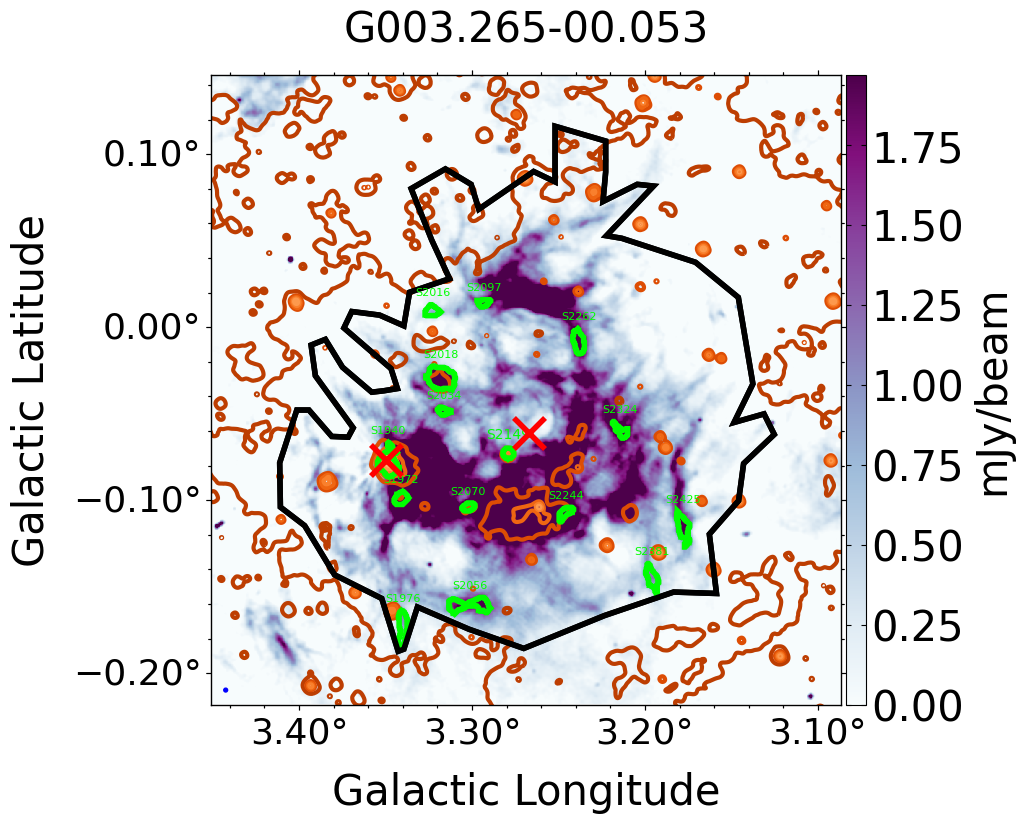}
    \label{fig:hii10}
  \end{subfigure}%
  \hspace{1em}%
  \begin{subfigure}{0.3\textwidth}
    \includegraphics[width=\linewidth]{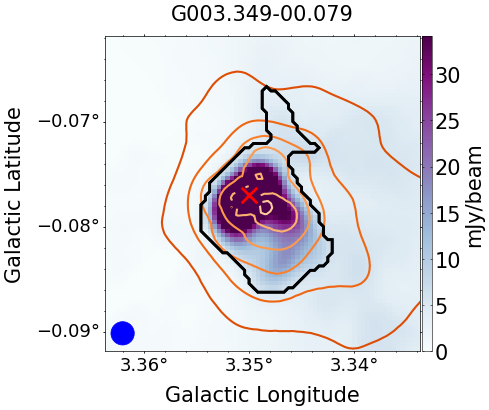}
    \label{fig:hii11}
  \end{subfigure}%
  \hspace{1em}%
  \begin{subfigure}{0.3\textwidth}
    \includegraphics[width=\linewidth]{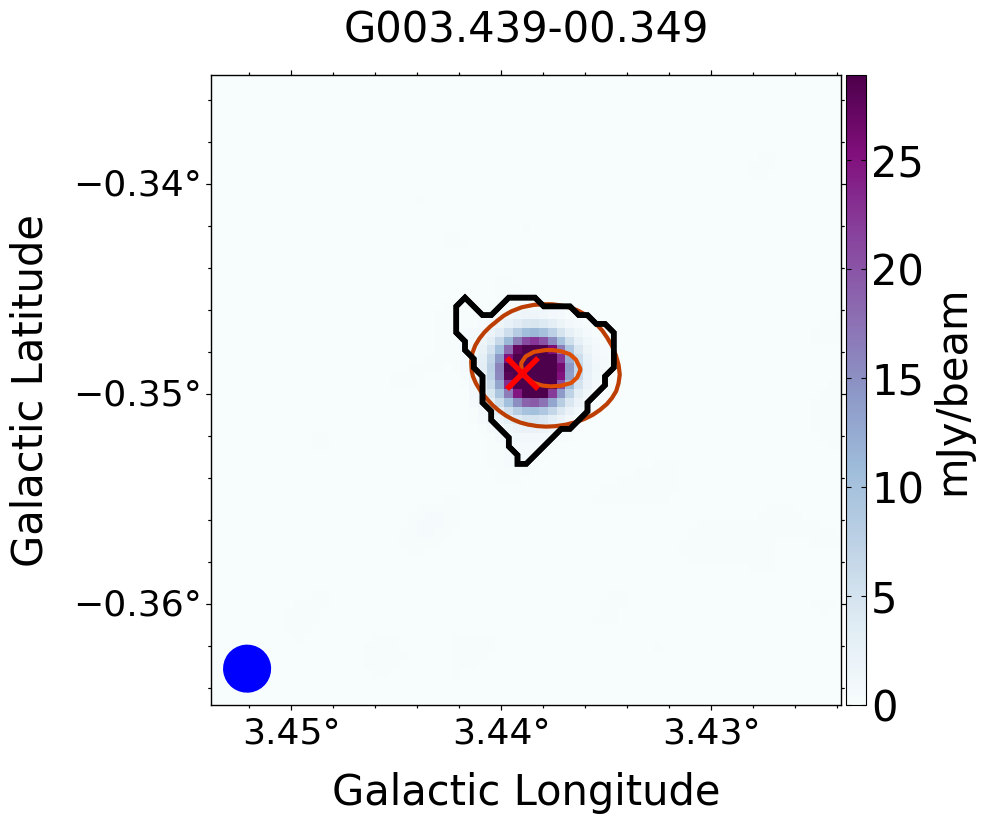}
    \label{fig:hii12}
  \end{subfigure}%
  \hspace{1em}%

  \caption{A sample of 1.3\ GHz SMGPS images and 12\ $\micro$m WISE known {$\textrm{H}\scriptstyle\mathrm{II}$} regions \rev{contours with 12 linearly spaced levels spanning the intensity range from the background near-zero level to the peak intensity}. The extent of the detected radio emission is indicated by the outer boundary, with sub-island substructures highlighted in different colours. The coordinates of the WISE {$\textrm{H}\scriptstyle\mathrm{II}$} region are marked with a red \rev{`}x' sign.  The beam size is shown as a solid circle in the lower-left corner of each panel. \rev{The panels are ordered by increasing Galactic longitude and represent a diverse cross-section of morphologies, from isolated, compact sources to complex, multi-scale structures.}}
  \label{fig:smultiplot}
\end{figure*}

\section{Physical Characterisation}
\label{results}

\subsection{Distance Measurements}
\label{results.1}

In order to determine kinematic distances for our sample\rev{,} we adopted the Galactic rotation model, derived from the Bar and Spiral Structure Legacy (BeSSeL) Survey \citep{2016ApJ...823...77R}, which is based on trigonometric parallax measurements of maser sources across the Milky Way. The kinematic distances to our target {$\textrm{H}\scriptstyle\mathrm{II}$} regions were computed using the Bayesian distance calculator of \citet{2019ApJ...885..131R}. Although the WISE catalogue of {$\textrm{H}\scriptstyle\mathrm{II}$} regions \citep{2014ApJS..212....1A} provides a comprehensive census of targets, it does not contain distance estimates for all sources in our sample. Furthermore, the distances reported in the WISE catalogue are largely compiled from the literature with a heterogeneous set of methods \citep{2014ApJS..212....1A}. To ensure internal consistency, we calculate the kinematic distances for our entire sample using a uniform technique. This method incorporates the best-fitting parameters for Galactic rotation and solar motion, while also accounting for the vertical distribution of massive young stars and the warp and precession of the Galactic plane \footnote{\url{https://www3.mpifr-bonn.mpg.de/staff/abrunthaler/bessel_calc2.0/}}. According to \citet{2019ApJ...885..131R}, the best-fitting Galactic rotation model indicates a rotation speed of approximately 237 km s$^{-1}$ at a Galactocentric radius of 6.8 kpc, gradually declining to about 227 km s$^{-1}$ at 14.1 kpc. The Bayesian formalism combines kinematic information with prior knowledge of the Galactic structure to yield the most likely heliocentric distance for each source, which we subsequently use to derive their physical sizes, Lyman-photon fluxes, and electron densities.

\begin{figure}
    \centering
    \includegraphics[width=\columnwidth]{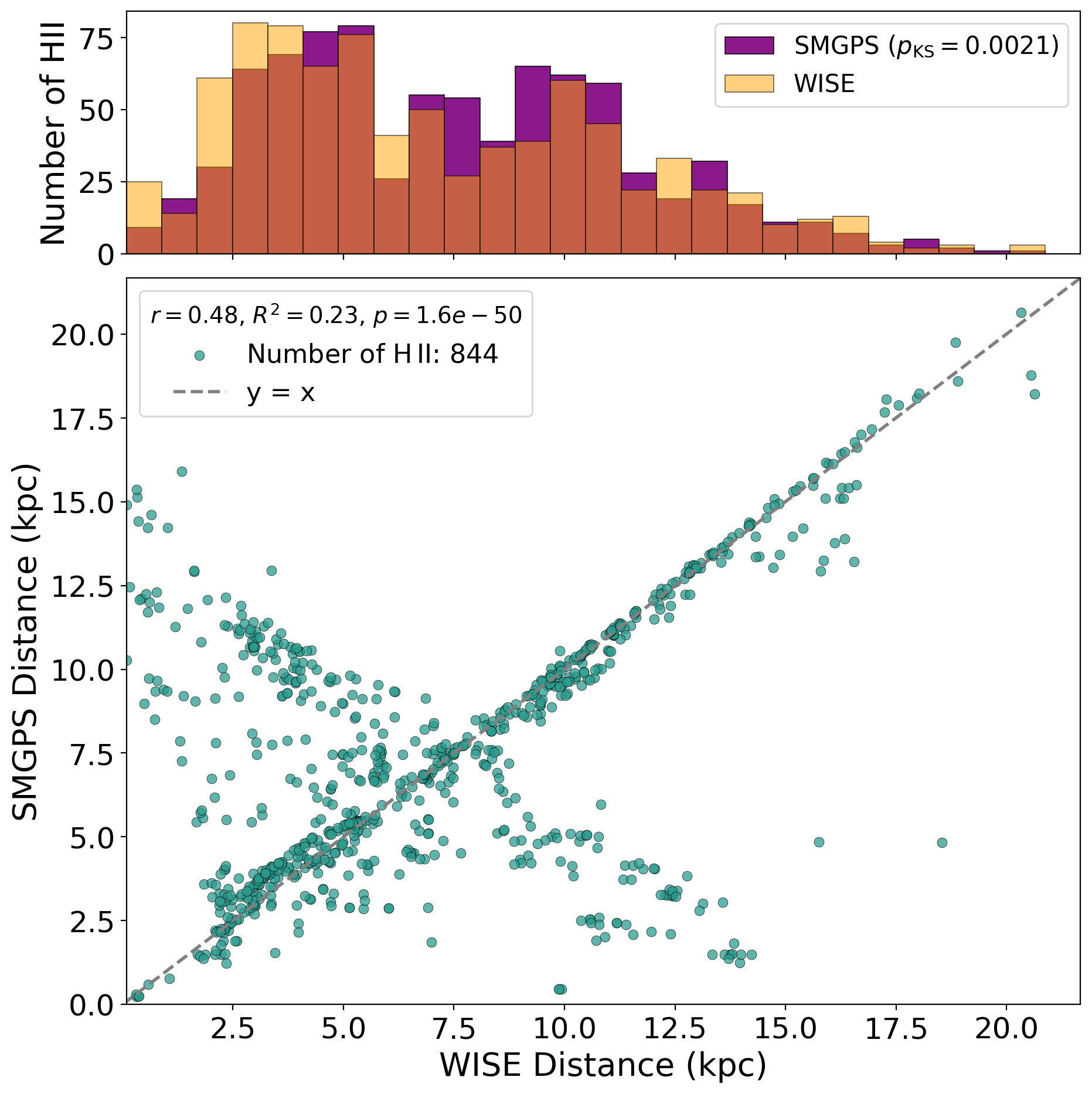}
    \caption{\rev{Comparison of heliocentric distance estimates for the matched \hii~regions with distance measurements ($N=844$). 
\textbf{Top:} Histograms of the heliocentric distance distributions for both datasets. While the samples trace the same large-scale Galactic structure, a KS-test ($p = 0.002$) suggests subtle distributional differences likely arising from the different KDA-resolution priors used.
\textbf{Bottom:} Scatter plot comparing distances derived through our Bayesian methodology ($d_{\rm SMGPS}$) against the WISE \hii catalog values ($d_{\rm WISE}$). The dashed black line represents the 1:1 identity. The presence of two distinct outlier populations (up to $\sim$15~kpc discrepancy) highlights the discrete nature of the near-far distance ambiguity. Correlation statistics (Pearson $r~0.483$, $R^2$, and $p < 0.001$) are provided in the legend.}}
    \label{fig:distances}
\end{figure}

To assess the reliability of our distance estimates, we compared them with those listed within the WISE H~{\sc ii} region catalogue. \rev{Figure~\ref{fig:distances} (top) presents the histograms of the heliocentric distance distributions for both datasets. While both samples trace the same large-scale Galactic structure, a Kolmogorov-Smirnov (KS) test yields a $p$-value of 0.002. This difference in distribution shapes likely arises from the specific spiral arm priors and distance-determination heuristics used in our Bayesian methodology compared to the heterogeneous methods in the WISE catalogue.} \rev{The lack of sources near $\sim8$~kpc reflects the survey footprint and the exclusion of the Galactic Centre ($|l| < 2^{\circ}$), a region observed independently \rev{with MeerKAT} by \citet{2022ApJ...925..165H}. This exclusion is a direct result of the longitude ranges selected for the SMGPS, which include $251^{\circ} \le l \le 358^{\circ}$ and $2^{\circ} \le l \le 61^{\circ}$.} \rev{The direct comparison between $d_{\rm SMGPS}$ and $d_{\rm WISE}$ is shown in the scatter plot in Figure~\ref{fig:distances} (bottom), showing a global Pearson correlation coefficient of $r = 0.483$ ($p < 0.001$).}

\begin{figure}
    \centering
    \includegraphics[width=\columnwidth]{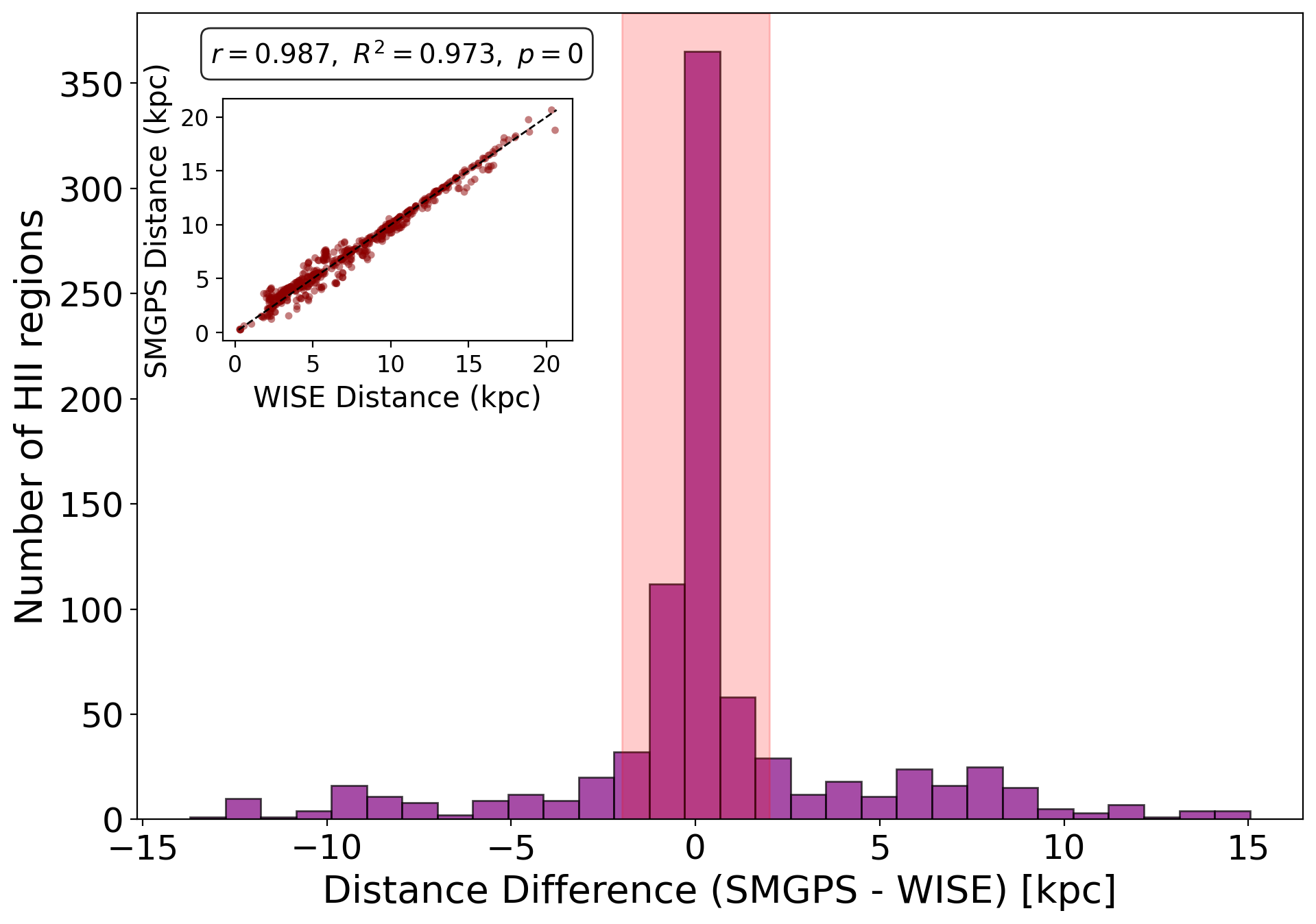}
    \caption{\rev{Statistical comparison of heliocentric distances between the SMGPS and WISE datasets. 
\textbf{Main panel:} Distribution of the distance residuals ($\Delta d = d_{\rm SMGPS} - d_{\rm WISE}$). The light shaded region highlights the high-agreement zone within $\pm 2$~kpc, which contains 68\% of the matched sample. The narrow central peak demonstrates that for the majority of regions, the Bayesian methodology is consistent with literature values. The broader, symmetric wings and distant outliers ($|\Delta d| > 5$~kpc) are characteristic of discrepancies in resolving the near-far kinematic distance ambiguity. 
\textbf{Inset:} Correlation between SMGPS and WISE distances for the subset within the high-agreement zone ($|\Delta d| \leq 2$~kpc). The high correlation coefficient ($r = 0.987$) and $R^{2}$ value indicate that when the distance ambiguity is resolved consistently, the two results are nearly identical.}}

    \label{fig:distance_diff_absolute}
\end{figure}

Figure~\ref{fig:distance_diff_absolute} shows the distribution of the kinematic distance differences between the SMGPS and WISE \hii~region sample. \rev{The distribution is strongly peaked around zero, with 68\% of the matched sample falling within a high-agreement zone of $\pm 2$~kpc (shaded red region). For this subset where the distance ambiguity is resolved consistently, the correlation is excellent ($r = 0.987, R^{2} = 0.973$; see Figure~\ref{fig:distance_diff_absolute} inset).} The residual scatter is expected, as the WISE distances were compiled from heterogeneous methods employed by different authors, including kinematic solutions, {$\textrm{H}\scriptstyle\mathrm{I}$} self-absorption techniques, and parallax measurements. The handful of large discrepancies (up to 15~kpc) most likely arise from disagreements in the near--far distance ambiguity for individual sources \rev{(KDA; e.g., \citealt{2012MNRAS.420.1656U, 2018ApJ...856...52W})}. While this will result in large differences in the distance-dependent properties of individual \hii~regions, it will not affect the statistical properties of the sample as a whole, as Figure~\ref{fig:distance_diff_absolute} shows.

\subsection{Radius Distribution}
\label{results.2}

\rev{Using the kinematic distances obtained in Section~\ref{results.1}, we calculate the physical radius of each H~{\sc ii} region as $R_{\mathrm{phys}}=D \tan\theta$, where $D$ is the distance and $\theta$ is the angular radius.} \rev{To assess the consistency between the radio and infrared observations without the inherent bias introduced by shared distance measurements, we compare the angular radii from the SMGPS \citep{2025A&A...695A.144B} and WISE \citep{2024ApJ...969...43A} catalogues (Figure~\ref{fig:radii}).}

\rev{The bottom panel of Figure~\ref{fig:radii} illustrates a direct linear relationship between the angular sizes, with the dashed line representing perfect correspondence ($y = x$). To rigorously quantify this relationship while accounting for any underlying distance dependence, we performed a partial Spearman correlation test. We find a strong intrinsic correlation of $\rho_{\mathrm{p}} = 0.84$ ($p \ll 0.001$), confirming that the two tracers are spatially co-extensive.} \rev{The best-fit power law (green solid line) yields a slope of $m = 1.15 \pm 0.02$. While slightly steeper than unity, the fit demonstrates that the radio and infrared sizes remain well-coupled across two orders of magnitude in angular scale.} \rev{Furthermore, the ratio of the angular radii ($\theta_{\mathrm{SMGPS}}/\theta_{\mathrm{WISE}}$) remains stable as a function of heliocentric distance (Figure~\ref{fig:radii}, top). The lack of a significant trend with distance suggests that the sensitivity and resolution of the SMGPS are sufficient to recover the extended morphology of these regions throughout the surveyed volume.} 

\rev{At smaller angular scales ($\theta \lesssim 1'$), we observe that the mid-infrared radii from the WISE catalogue are systematically larger than the 1.3~GHz radio radii. Rather than a measurement error, this likely reflects a physical difference in the emission zones: the 12~$\mu$m emission in WISE traces the surrounding PAH emission in the Photo-Dissociation Region (PDR), whereas the SMGPS radio continuum specifically traces the interior volume of ionised gas \citep{2014ApJS..212....1A}.} \rev{Furthermore, the discrepancies may be influenced by the different source characterization methods, with WISE radii defined by the visual extent of the infrared envelope, while the SMGPS uses the automated {\tt{CAESAR}} algorithm to fit the radio contours.}

\begin{table*}
\centering
\small
\setlength{\tabcolsep}{4pt}
\renewcommand{\arraystretch}{1.2}
\caption{Physical properties of SMGPS {$\textrm{H}\scriptstyle\mathrm{II}$} regions,
 including integrated 1.3\ GHz flux densities,
 mean line-of-sight velocities ($v_\mathrm{LSR}$),
 and estimated ionising photon flux densities. The distances
 were estimated using \citet{2019ApJ...885..131R} Bayesian calculator and the
 stellar spectral classifications are inferred
 following the calibrations of \citet{1973AJ.....78..929P}.}
\label{tab:hii_sample}
\begin{adjustbox}{width=\textwidth, center}
\begin{tabular}{llccccccccc}
\toprule
SMGPS Name & WISE Name & $F_{12\mu\text{m}}$ & $v_{\text{LSR}}$ & $D \pm \delta D$ & $R_{\text{WISE}}$ & $F_{\text{radio}} \pm \delta F$ & $R_{\text{radio}}$ & $n_{\text{e}}$ & $\log N_{\text{Ly}}$ & Spec. \\
 & & (Jy) & (km/s) & (kpc) & (pc) & (Jy) & (pc) & (cm$^{-3}$) & (s$^{-1}$) & Type \\
\midrule
G001.324+00.102 & G001.321+00.101 & -- & -12.7 & $3.7 \pm 0.2$ & 1.1 & $3.7e-01 \pm 5.4e-03$ & 1.3 & 25.9 & 47.7 & O9.5 \\
G001.328+00.151 & G001.330+00.145 & -- & -20.1 & $3.7 \pm 0.2$ & 1.4 & $5.2e-02 \pm 1.5e-03$ & 0.6 & 31.5 & 46.8 & B0 \\
G001.331+00.085 & G001.330+00.088 & -- & -12.0 & $3.7 \pm 0.2$ & 0.8 & $1.7e-01 \pm 4.3e-03$ & 0.6 & 54.4 & 47.4 & B0 \\
G001.487-00.200 & G001.488-00.199 & -- & -2.2 & $12.6 \pm 0.3$ & 2.6 & $5.2e-02 \pm 7.9e-04$ & 2.1 & 15.4 & 47.9 & O9 \\
G002.011-00.680 & G002.009-00.680 & -- & 18.2 & $2.8 \pm 0.2$ & 0.6 & $1.6e-01 \pm 2.4e-04$ & 1.1 & 15.5 & 47.1 & B0 \\
G002.273+00.236 & G002.272+00.237 & -- & 4.9 & $2.8 \pm 0.2$ & 3.9 & $3.6e+00 \pm 2.0e-02$ & 4.6 & 8.9 & 48.4 & O7.5 \\
G002.417-00.610 & G002.422-00.611 & -- & 7.8 & $2.8 \pm 0.2$ & 1.2 & $4.5e-02 \pm 7.2e-04$ & 0.8 & 13.7 & 46.5 & B0 \\
G002.528-00.035 & G002.521-00.024 & -- & 8.3 & $2.8 \pm 0.2$ & 2.3 & $7.9e-01 \pm 4.4e-03$ & 5.3 & 3.3 & 47.8 & O9.5 \\
G002.604+00.141 & G002.614+00.133 & -- & 102.4 & $8.1 \pm 2.2$ & 2.4 & $4.4e-01 \pm 2.4e-03$ & 3.3 & 14.4 & 48.4 & O7.5 \\
G003.265-00.053 & G003.266-00.061 & -- & 7.1 & $12.6 \pm 0.2$ & 23.9 & $1.0e+01 \pm 6.5e-03$ & 45.8 & 2.1 & 50.2 & O4 \\
G003.349-00.079 & G003.349-00.078 & -- & 8.3 & $12.6 \pm 0.2$ & 4.4 & $6.4e-01 \pm 4.6e-03$ & 2.6 & 38.8 & 49.0 & O6 \\
G003.439-00.349 & G003.439-00.349 & -- & -21.6 & $4.7 \pm 0.2$ & 1.0 & $7.6e-02 \pm 7.6e-05$ & 0.5 & 65.2 & 47.2 & B0 \\
G003.442-00.643 & G003.442-00.650 & -- & 0.1 & $1.5 \pm 0.1$ & 0.9 & $1.4e-01 \pm 2.8e-04$ & 1.1 & 7.9 & 46.5 & B0 \\
G003.649-00.119 & G003.650-00.122 & -- & 4.9 & $2.9 \pm 0.2$ & 1.0 & $3.9e-01 \pm 1.9e-04$ & 1.2 & 22.8 & 47.5 & O9.5 \\
G003.665-00.099 & G003.664-00.098 & -- & 6.2 & $12.6 \pm 0.2$ & 1.9 & $9.3e-02 \pm 5.2e-04$ & 1.9 & 22.9 & 48.1 & O8.5 \\
G003.930-00.117 & G003.928-00.116 & -- & 32.1 & $12.6 \pm 0.3$ & 2.6 & $1.5e-01 \pm 2.6e-04$ & 3.1 & 14.1 & 48.4 & O7.5 \\
G004.350+00.155 & G004.343+00.115 & -- & 7.0 & $2.9 \pm 0.2$ & 1.6 & $1.3e+00 \pm 4.2e-03$ & 4.1 & 6.4 & 48.0 & O9 \\
G004.435+00.069 & G004.409+00.110 & -- & 4.1 & $2.9 \pm 0.2$ & 3.4 & $5.7e+00 \pm 4.7e-03$ & 5.6 & 8.7 & 48.7 & O6.5 \\
G004.565-00.118 & G004.557-00.124 & -- & 10.2 & $2.9 \pm 0.2$ & 2.2 & $7.4e-01 \pm 7.6e-03$ & 2.4 & 11.2 & 47.8 & O9.5 \\
G005.064+00.281 & G005.068+00.275 & -- & 12.0 & $2.9 \pm 0.2$ & 3.9 & $1.4e+00 \pm 1.5e-02$ & 7.8 & 2.6 & 48.1 & O9 \\

\bottomrule
\end{tabular}
\end{adjustbox}
\begin{flushleft}
\footnotesize \textbf{Note:} The complete version of this table is available in machine-readable format in the online journal. Appendix Table~\ref{tab:source_properties} \rev{provides} full column descriptions.
\end{flushleft}
\end{table*}

\subsection{Ionising Photon Flux and Electron Density}

We exploit the exceptional sensitivity of SMGPS observations to delineate the morphology of {$\textrm{H}\scriptstyle\mathrm{II}$} regions within our target sample. For many of the {$\textrm{H}\scriptstyle\mathrm{II}$} regions in the sample, the SMGPS represents their first observation at radio wavelengths with an angular resolution $< 45$ arcsec. The high-resolution flux density measurements obtained from SMGPS data allow a more precise determination of the spatial distribution of the ionised gas within these regions. This enhanced characterisation is crucial for accurately estimating the ionising photon fluxes emanating from the embedded young stellar objects that power these Galactic {$\textrm{H}\scriptstyle\mathrm{II}$} regions.

The $\SI{1.3}{\giga\hertz}$ radio continuum fluxes toward the 1,327 objects in our catalogue are presented in Table \ref{tab:hii_sample}. The reported flux density values and their uncertainties were extracted from the SMGPS extended source catalogue \citep{2025A&A...695A.144B}, where the total uncertainty is calculated as the quadrature sum of the statistical uncertainty and a $\mathbf{5\%}$ calibration uncertainty. To determine the spectral types of the candidate ionising stars, we first estimated the Lyman continuum photon flux ($N_{\text{Ly}}$), assuming that the radio emission is optically thin. This assumption is valid for "classical" {$\textrm{H}\scriptstyle\mathrm{II}$} regions \citep{2012AJ....144..173D}. We acknowledge that for optically thick sources, the resulting  $N_{\text{Ly}}$ value represents a lower limit as a result of self-absorption and photon absorption by dust. We calculate $N_{\text{Ly}}$ (in $\text{s}^{-1}$) following the formulation used by \citet{2012AJ....144..173D} (see also \citep{aa.30.090192.003043, 2018ARep...62..764T}):

\begin{figure}
    \centering
        \includegraphics[width=\columnwidth]{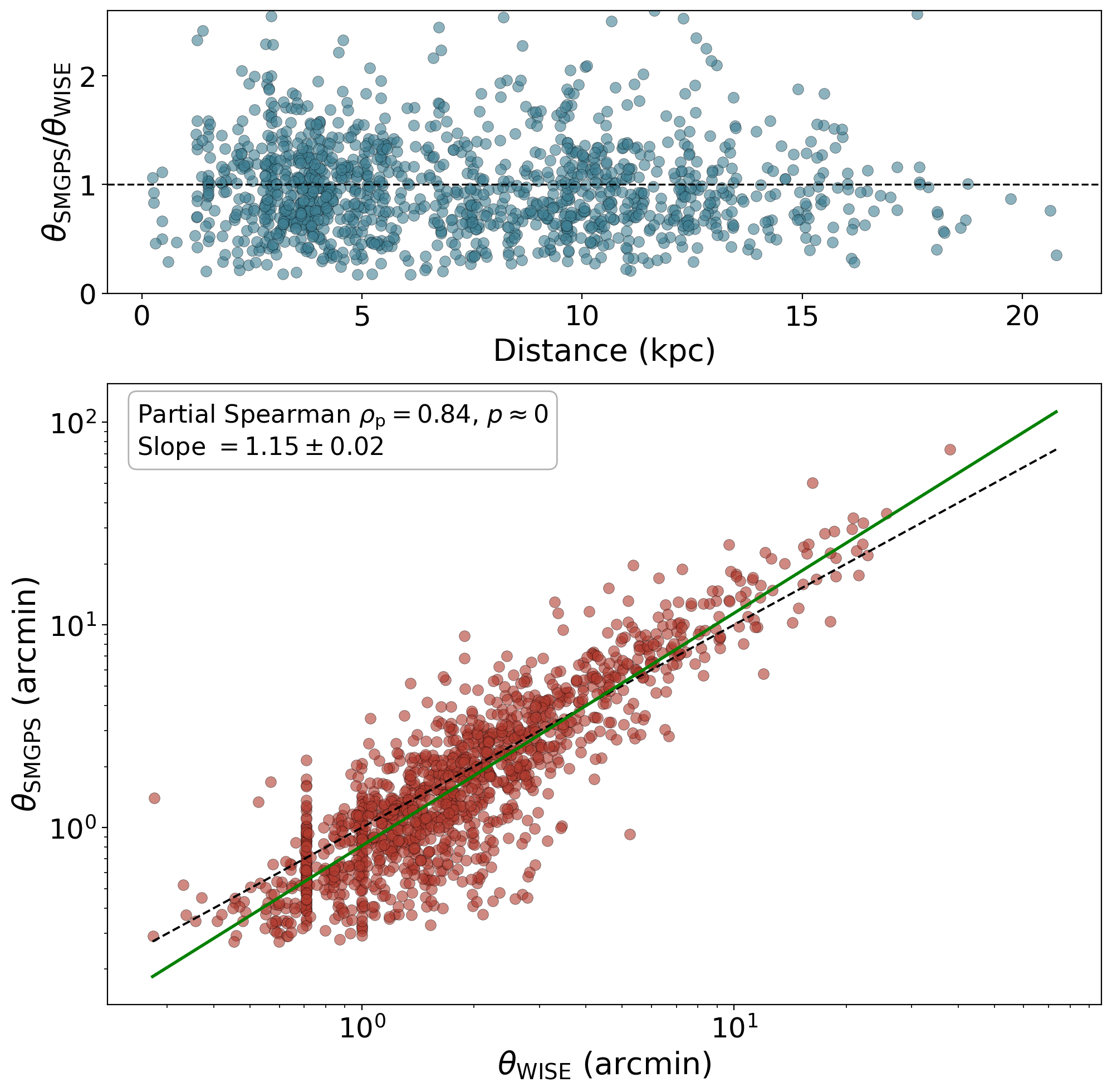}
    \caption{\rev{Comparison of \hii~region angular radii between SMGPS and WISE. 
    \textbf{Top:} Ratio of radio-to-infrared angular radii ($\theta_{\rm SMGPS}/\theta_{\rm WISE}$) as a function of heliocentric distance in kpc. The stability of the ratio around unity across the survey range suggests minimal distance-dependent bias in our size estimates. 
    \textbf{Bottom:} Direct correlation between angular radii. The dashed line indicates $y=x$, while the solid line shows a linear fit used to determine the scaling slope ($1.15\pm0.02$) between the two surveys. To account for the shared dependence on distance in physical radii, we provide the partial Spearman correlation ($\rho_{\rm p} = 0.84$), which confirms a strong intrinsic spatial correlation ($p < 0.001$) between the radio continuum and mid-infrared emission.}}

    \label{fig:radii}
\end{figure}

\rev{
\begin{equation}
\label{eq:nly}
N_{Ly}\approx7.54\times10^{46}\left(\frac{T_{\rm e}}{\SI{10000}{\kelvin}}\right)^{-0.45}
\left(\frac{\nu}{\SI{}{\giga\hertz}}\right)^{0.1}
\left(\frac{S_\nu}{\text{Jy}}\right)
\left(\frac{D}{\text{kpc}}\right)^2\;s^{-1}
\end{equation}
}

\noindent where $S_{\nu}$ is the observed specific flux density  (e.g. at 1.3~GHz), $D$ is the distance to the {$\textrm{H}\scriptstyle\mathrm{II}$}  region, $T_{\text{e}}$ is the electron temperature and $\nu$ is the observation frequency. Since some sources lack direct temperature measurements (see Figure \ref{fig:te} for available data from \rev{\citealt{2019ApJS..240...24W}}), we adopted a representative value of 6,200 K, determined from the average $T_{\rm e}$ measurement from \citet[][see Section 3]{2019ApJS..240...24W}. We note that, given the $T_{\rm e}^{-0.45}$ dependence of $N_{\rm Ly}$, an error in estimating the electron temperature does not result in an appreciable error in the derived \rev{ionising photon flux. For instance, considering our adopted mean electron temperature of $\sim6000$~K, an uncertainty of $\pm 2000$~K (a 33\% variation) results in a change in $N_{\rm Ly}$ of only $\sim$10\%. Since $N_{\rm Ly}$ spans several orders of magnitude, this minor dependency ensures our spectral classifications remain statistically robust.} For the kinematic distance estimation, we adopt the most likely values from the Bayesian distance calculator of \citet{2019ApJ...885..131R}. The spectral types of the candidate ionising stars were then inferred by comparing their derived $N_{\mathrm{Ly}}$ values with predictions from established stellar atmosphere models \citep{1973AJ.....78..929P}.

\begin{figure}
    \centering
    \includegraphics[width=\columnwidth]{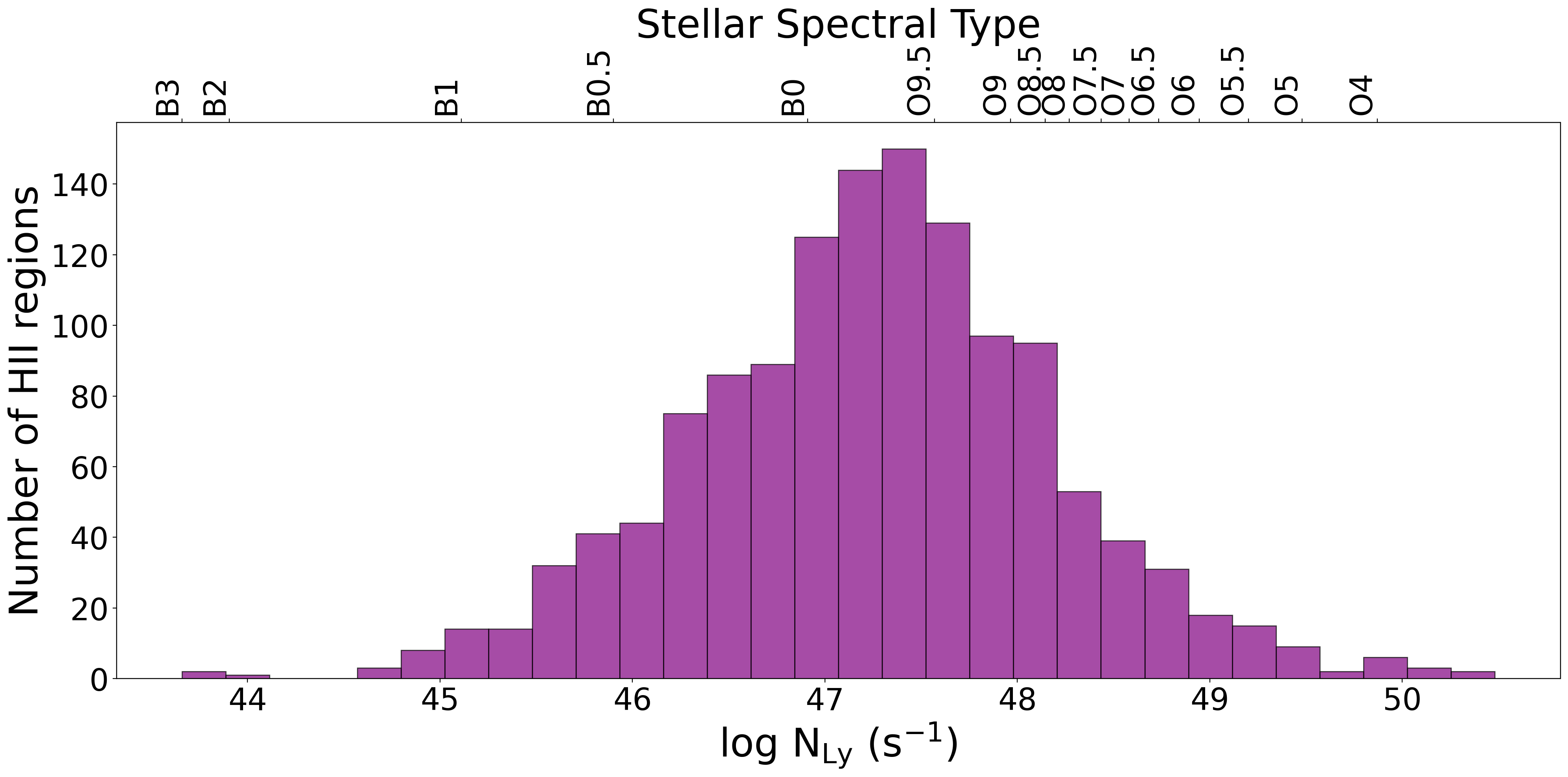}
    \caption{Distribution of ionising photon flux densities for 1,327 SMGPS {$\textrm{H}\scriptstyle\mathrm{II}$} regions. The histogram shows the log-scaled ionising photon flux (log $N_{\text{Ly}}$) with corresponding stellar spectral types, ranging from B2 to O4, indicated on the top axis. The distribution \rev{of ionising photon fluxes peaks at $\log N_{\rm Ly} \approx 48.18$, corresponding to the transition between O8 and O7.5 spectral types. This primary peak region constitutes approximately 15\% of the sample, indicating that these are the typical ionising sources for the classical \hii~regions in our survey.}}

    \label{fig:nly_histogram}
\end{figure}

The resulting spectral types range from B2 to O4. \rev{The ionising photon flux distribution peaks at $\log N_{\rm Ly} = 48.18 \rm~s^{-1}$, which corresponds to a late-O star (roughly O7.5 subclass on the \citealt{1973AJ.....78..929P} scale). These sources constitute approximately 15\% of our total sample, as shown in  Figure~\ref{fig:nly_histogram}
%with up 40\% representations, see Figure \ref{fig:nly_histogram}).
The results align with the expected population distribution of Galactic \hii~regions \citep{2019A&A...628A...6S}. However, it is important to acknowledge that this classification assumes the ionising flux is provided by a single central star. In reality, many classical \hii~regions are powered by clusters containing multiple massive stars. If a cluster is present, the single-star assumption tends to overestimate the mass of the primary ionising source. Following standard conventions (e.g., \citealt{2005A&A...436.1049M}), the most massive star in the cluster is likely one spectral sub-class later than the value derived from the total $N_{\rm Ly}$ flux.}

\begin{figure}
    \centering
    \includegraphics[width=\columnwidth]{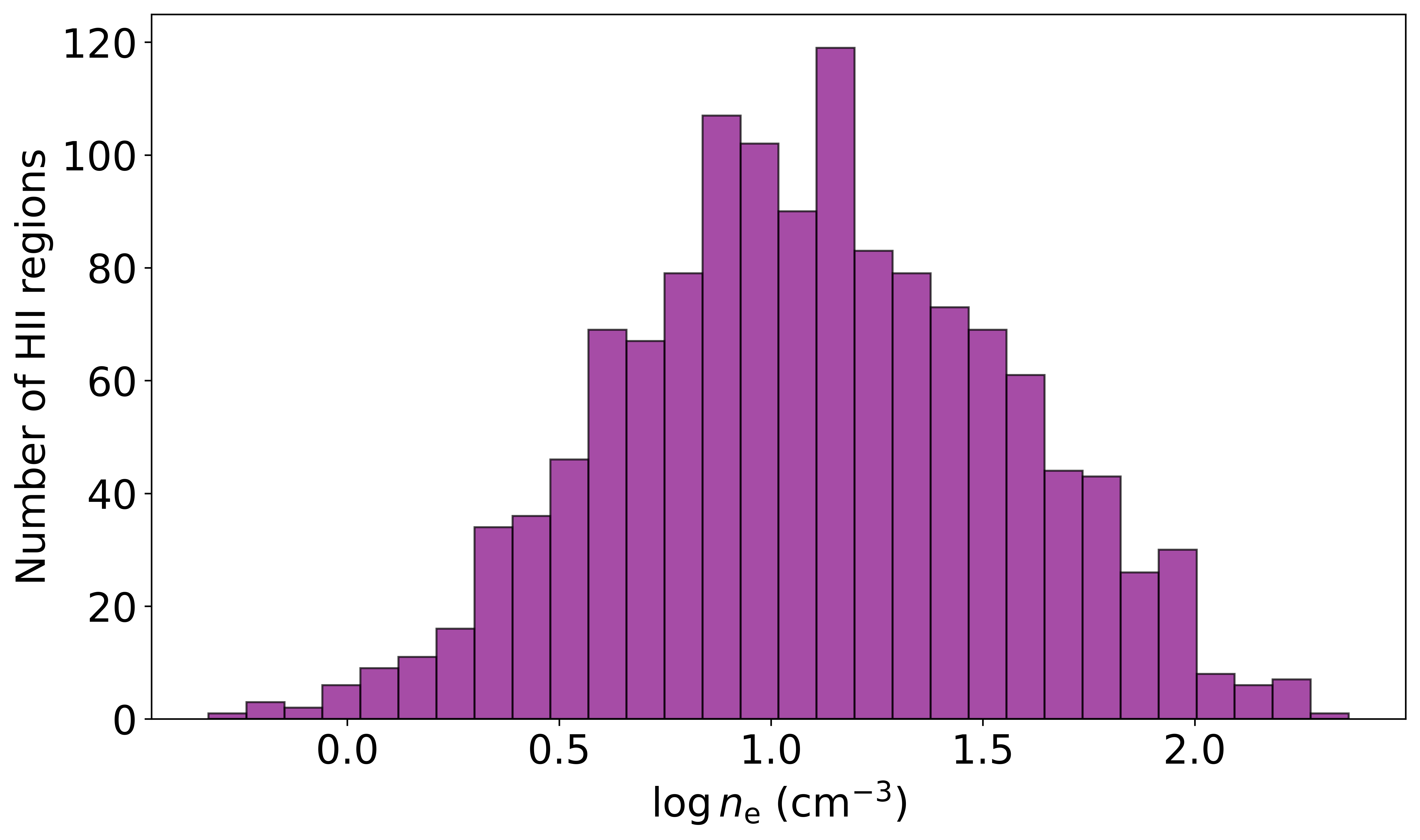}
    \caption{Electron density ($\rev{\log} n_{{\rm \rev{e}}}$) distribution for SMGPS \hii~region sample derived from RRL observations. The sources fall below the $n_\text{\rev{e}} < 10^4 \text{ cm}^{-3}$ threshold, which is characteristic of classical diffuse \hii~regions according to the classification by \citet{2002ARA&A..40...27C}.}
    \label{fig:ne_dist}
\end{figure}

\rev{To calculate the physical properties of the ionised gas, specifically the electron density ($n_{\text{e}}$ in $\text{cm}^{-3}$), we assume a spherical, homogeneous model. Following the standard derivation of the Str\"omgren condition \citep{1939ApJ....89..526S}, the relationship between the ionising photon flux ($N_{\rm Ly}$) and the physical parameters of the region can be expressed as a function of observed radio flux and distance:}

\begin{equation}
\label{eq:ne}
\begin{split}
n_{\rm e}
\simeq 7200 \, 
&\left( \frac{S_\nu}{\mathrm{Jy}} \right)^{1/2} 
\left( \frac{T_{\rm e}}{\mathrm{K}} \right)^{0.175} 
\left( \frac{\nu}{\mathrm{GHz}} \right)^{0.05} 
\left( \frac{\theta_\mathrm{}}{\mathrm{arcsec}} \right)^{-1} \\
&\times \left( \frac{R_\mathrm{}}{\mathrm{pc}} \right)^{-1/2} 
\left( \frac{\alpha_B}{2.6\times10^{-13}\,\mathrm{cm^3\,s^{-1}}} \right)^{-1/2} \mathrm{cm^{-3}},
\end{split}
\end{equation}

\noindent where $R$ is the physical radius of the \hii~region, $\theta$ is the angular size, and $\alpha_{\text{B}}$ is the Case $\text{B}$ recombination coefficient. \citet{2006agna.book.....O} show that for $T_\text{\rev{e}} \approx$ 6,200 K the effective $\alpha_{\text{B}}$ is $2.6 \times 10^{-13}$ cm$^{3}$s$^{-1}$. This calculation allows us to distinguish between the dense \hii~population and the more diffuse, evolved \hii~regions. The $n_\text{\rev{e}}$ distribution (Figure~\ref{fig:ne_dist}), shows that our sample falls below the $n_\text{\rev{e}} < 10^4 \text{ cm}^{-3}$ threshold, which according to the \citet{2002ARA&A..40...27C} classification is characteristic of classical \hii~regions.

\begin{figure}
    \centering
    \includegraphics[width=\columnwidth]{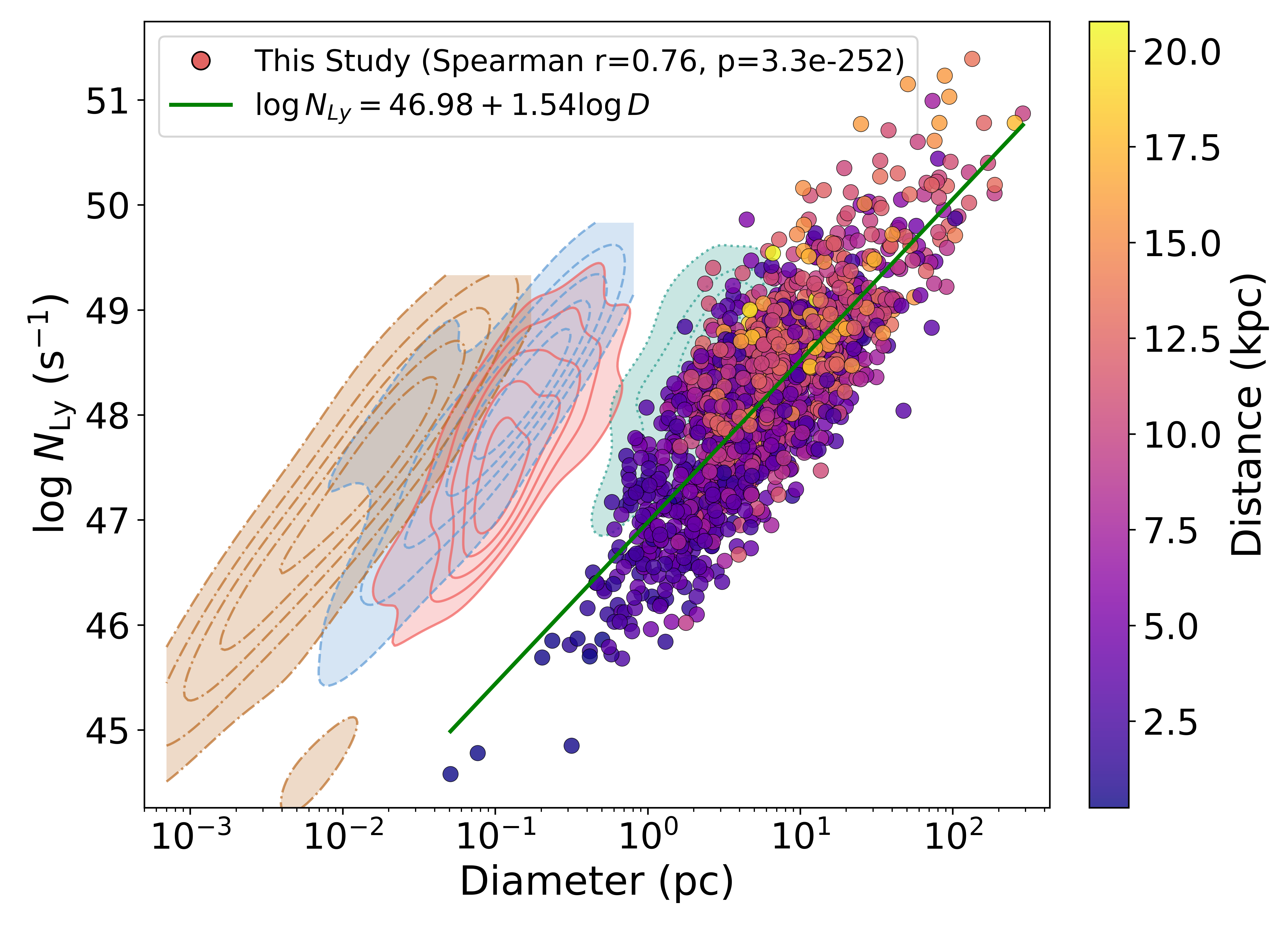}
    \caption{
    The ionising photon flux–size relation for SMGPS \hii~regions. The solid green line represents \rev{a least-squares power-law fit to the SMGPS data, with a slope of $1.54\pm0.03$ and a Spearman r = 0.76 (p $< 0.001$). Points are colour-coded by distance from the Sun in kpc}. \rev{The contours represent the density estimation of various \hii~region evolutionary stages from the literature, including hypercompact represented by dash-dot \citep{2025MNRAS.538.2267P} and dashed contours \citep{2021A&A...645A.110Y}, ultracompact in solid contours \citep{2018A&A...615A.103K}, and compact regions in dotted contours \citep{2024A&A...689A..81K}. The SMGPS sample effectively extends the census into the classical, diffuse regime, showing that the upper limit of ionising output remains relatively constant ($N_{\text{Ly}} \approx 10^{47}$–$10^{48}$ s$^{-1}$) across evolutionary stages. The solid black arrow represents the predicted evolutionary track of a compact \hii~region during its expansion phase.}}
    \label{fig:nly_size}
\end{figure}

\begin{figure}
    \centering
    \includegraphics[width=\columnwidth]{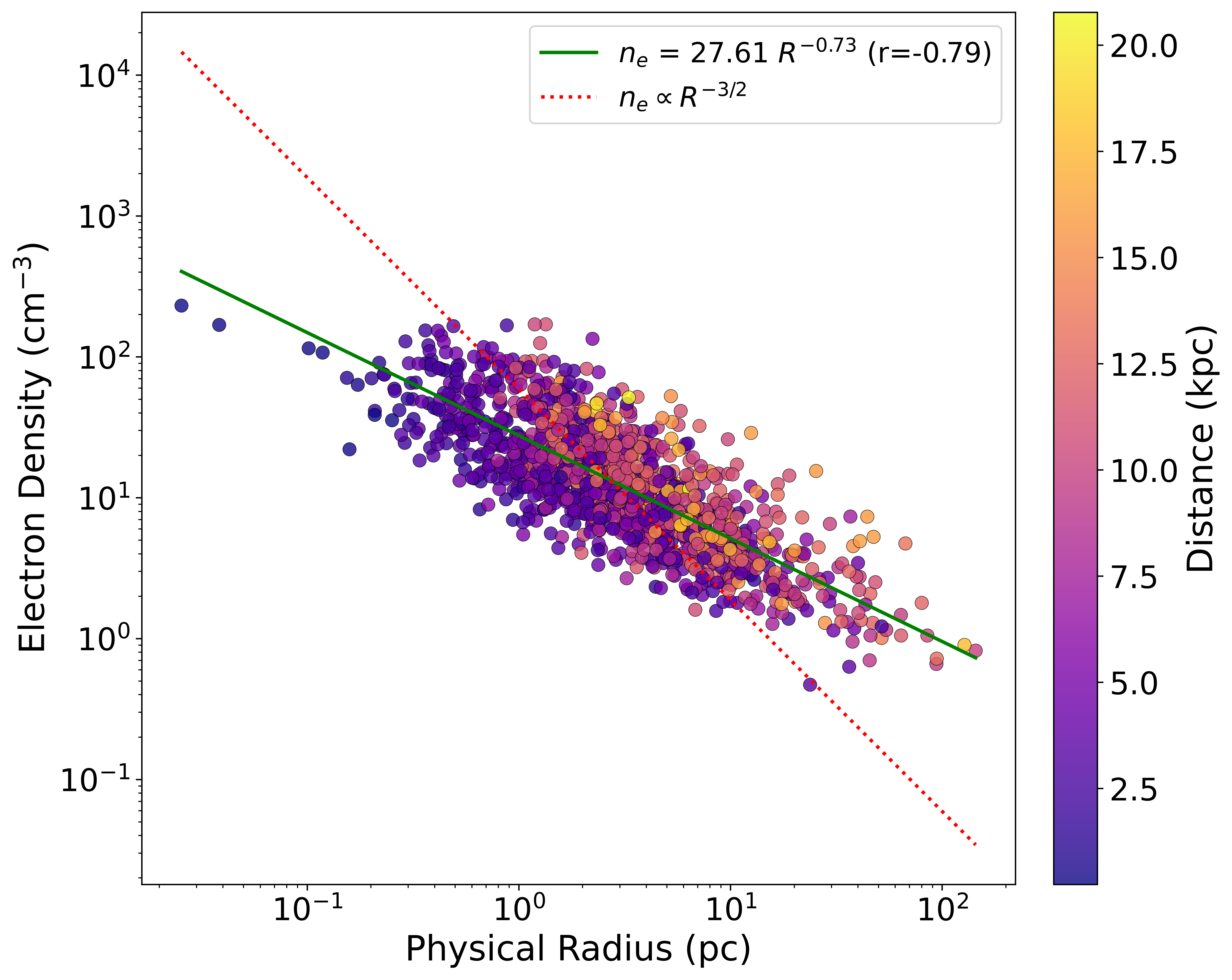}
    \caption{The electron density–size relation for SMGPS line shows the R$^{-3/2}$ relation expected for a spherical, uniform-density nebula; the solid line is a least-squares regression to the data points \rev{with a gradient of $-0.73\pm0.27$ (r=0.79)}. Points are colour-coded by distance from the Sun in kpc.}
    \label{fig:size_ne}
\end{figure}

\begin{figure}
    \centering
    \includegraphics[width=\columnwidth]{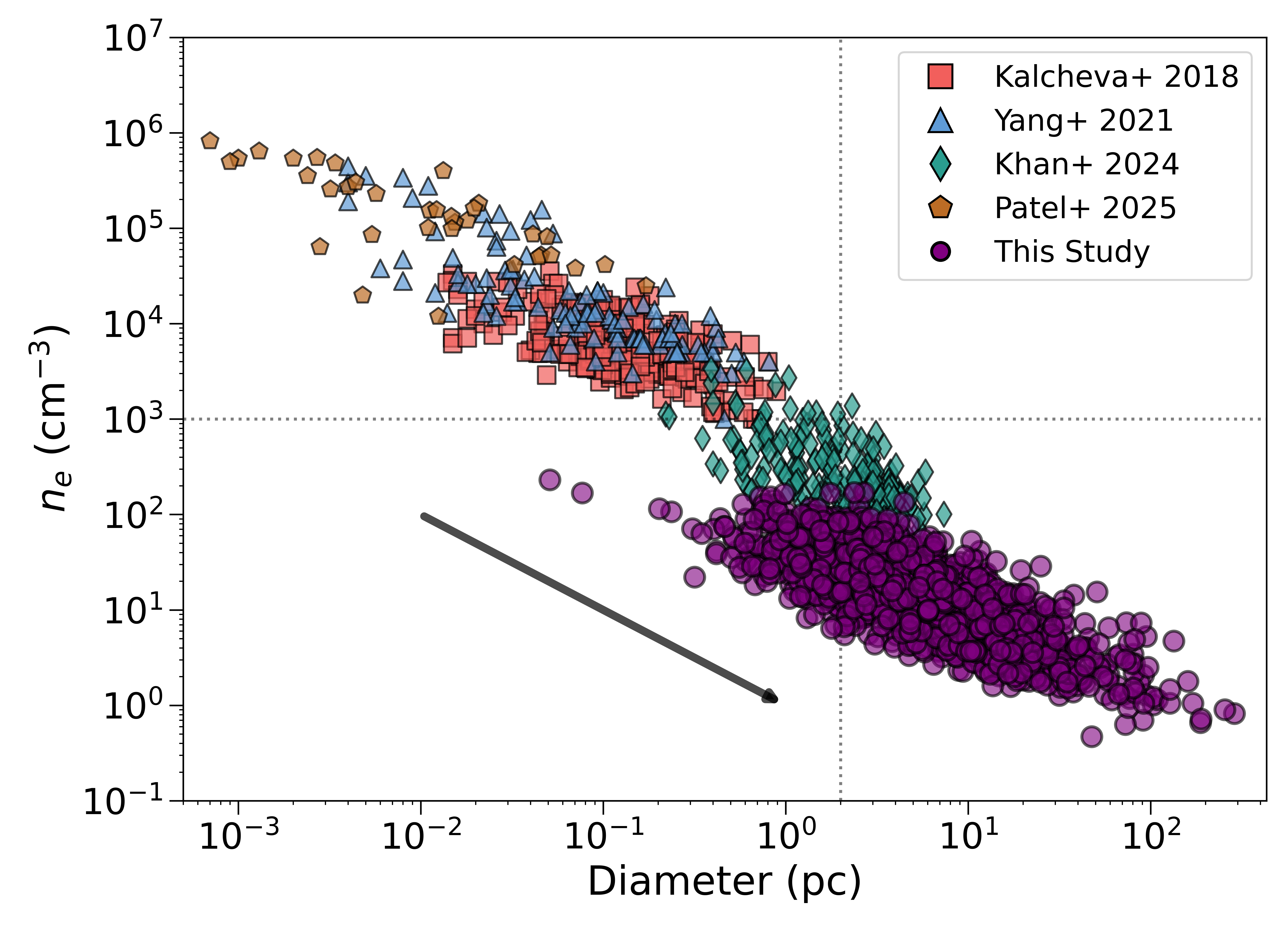}
    \caption{\rev{Comparison of the electron density–size relation between the SMGPS sample (circles) and various literature surveys. Data points include C\hii~regions from \citet{2024A&A...689A..81K} (diamonds), UC\hii~regions from \citet{2018A&A...615A.103K} (squares) and HC\hii~ from \citep{2021A&A...645A.110Y} (triangles) and \citet{2025MNRAS.538.2267P} (pentagons). The arrow illustrates the characteristic evolutionary path of a compact \hii~region, representing the typical trajectory of these objects as they expand. The SMGPS data provide a continuous transition from compact to classical diffuse regions ($D > 1$ pc), confirming a consistent evolutionary trend across five orders of magnitude in physical diameter. The horizontal and vertical dotted lines represent the generally accepted upper limits for electron density ($n_{\text{e}} \approx 10^3 \text{ cm}^{-3}$) and physical diameter ($D \approx 2$ pc) for the classical \hii~region regime, respectively \citep{2002ARA&A..40...27C}.}}
    \label{fig:size_ne_lit}
\end{figure}

\section{Discussion}
\label{discussion}

\rev{The analysis of the SMGPS sample provides a high-sensitivity census of evolved \hii~regions in the Galactic plane. Our results demonstrate a remarkably consistent relationship between the 1.3~GHz radio continuum and the 12~$\mu$m mid-infrared emission across a wide range of angular scales. The direct comparison of angular radii ($\theta$) reveals an intrinsic correlation of $\rho_{\rm p} = 0.84$ after controlling for distance, confirming that the ionised gas and the surrounding dust envelopes are spatially co-extensive. Furthermore, the electron density distribution ($n_{\rm e} < 10^4 \text{ cm}^{-3}$) confirms that the majority of our sources have evolved out of the ultra-compact phase and are now in the diffuse, classical stage of expansion.}

\subsection{Sample Properties}
\label{sec:properties}

Before dwelling on the properties of the sample as a whole, it is instructive to first consider the sample completeness and any inherent selection biases that are present in the way in which we have identified these \hii~regions. Our sample is selected from those objects in the WISE \hii~region catalogue that are detected at 1.3 GHz by the SMGPS and which have existing radio recombination line (RRL) measurements.
\rev{As demonstrated by \citet{2024MNRAS.531..649G}, the SMGPS sensitivity limits are sufficient to detect an isothermal, optically thin, homogeneous H~{\sc ii} region powered by a B0.5 star up to a distance of 18~kpc (assuming standard scaling relations, e.g., \citealt{1990ApJ...362..147C, 2003ApJ...599.1333S}). This baseline capability confirms that the SMGPS data itself could, in theory, recover these fainter, distant sources. Given the overall SMGPS sensitivity ($\sigma \approx 15\,\mu\text{Jy/beam}$), our current sample is primarily limited by the requirement of a corresponding RRL detection for distance determination.}
The RRL measurements provided in the \citet{2014ApJS..212....1A} WISE \hii\ region catalogue are drawn from the literature \citep[mainly single-dish observations, e.g.][]{1987A&A...171..261C, 2010ApJ...718L.106B} plus dedicated interferometric follow-up \citep[e.g.]{2019ApJS..240...24W, 2021ApJS..254...36W,2011ApJS..194...32A,2018ApJS..234...33A}. Consequently, our sample of \hii\ regions is likely biased towards bright and extended continuum sources that satisfy the detection requirements for RRLs. As mentioned in Section \ref{catalogue}, there are many more (\rev{$\sim$ 1500}) SMGPS-associated \hii\ region candidates without RRL detections than with. This will have an effect on the completeness of our sample, which is discussed further in Section \ref{galtrends}.

We \rev{compare the ionising photon flux–size relation of the SMGPS sample to various \hii~region evolutionary stages from the literature, including hypercompact \citep{2021A&A...645A.110Y, 2025MNRAS.538.2267P}, ultracompact \citep{2018A&A...615A.103K}, and compact \citep{2024A&A...689A..81K} \hii~regions (see Figure \ref{fig:nly_size}). By populating the previously undersampled regime of more evolved, classical \hii~regions, the SMGPS data confirms that the upper limit of ionising output remains relatively constant ($N_{\text{Ly}} \approx 10^{47}$–$10^{48}$ s$^{-1}$) across evolutionary stages. This is consistent with the predicted evolutionary track of expanding \hii~regions (indicated by the black arrow in Figure  \ref{fig:nly_size}) and aligns with the expected constant Lyman flux of massive stars during their main-sequence lifetime.} The greater sensitivity of SMGPS means that we are able to populate the region of the N$_{\rm Ly}$--diameter space that is missing from the \citet{2025MNRAS.538.2267P} diagram and confirm that this trend continues into the classical \hii~region regime.

Figure \ref{fig:size_ne} presents the electron density–size relation for the sample of the SMGPS $\textrm{H}\scriptstyle\mathrm{II}$ regions, revealing a clear inverse correlation between electron density and physical radius, with a least-squares fit that yeilds $n_{\mathrm{e}} = 27.61R^{-0.73}$, with a moderate correlation coefficient of $r=-0.79$. 
This \rev{inverse-proportional relationship between electron density and source diameter closely mirrors the trends reported in foundational studies \citep{2001ApJ...549..979K} as well as more recent high-resolution surveys of younger \hii~region populations \citep{2025MNRAS.538.2267P}}. The observed slope is significantly shallower than the $R^{-3/2}$ dependence expected for classical Str\"omgren spheres expanding into a uniform, homogeneous medium. As discussed by \citet{2001ApJ...549..979K}, this discrepancy is naturally explained if more compact \hii~regions form within systematically denser environments, such as the inner regions of molecular cloud cores. The continuity of the density–physical size relation therefore suggests that the SMGPS \hii~regions sampled a range of evolutionary stages, rather than representing distinct, disconnected populations. This is consistent with theoretical expectations of \hii~region expansion, in which young, compact regions are confined by high ambient densities before expanding into progressively lower-density surroundings. \rev{The smooth transition between the literature samples (representing UC\hii~and C\hii~regions) and the SMGPS sample (representing classical regions) underscores the evolutionary continuity probed by this survey (see Figure \ref{fig:size_ne_lit}).}

%\newpage

\subsection{Radio–Infrared Relations}
%{$\textrm{H}\scriptstyle\mathrm{II}~$}
The physical synergy between the ionised gas and its surrounding dust environment is further reflected in the tight correlation between the radio and infrared fluxes. Studies such as those performed by \citet{2018ARep...62..764T} have shown a linear relationship between radio continuum fluxes (which are proportional to ionising photon flux) and infrared fluxes at various wavelengths (e.g., 8 and 24 $\micro$m). \citet{2014ApJS..212....1A} observed a similar relation but between 24 $\mu$m and 21 cm fluxes. Studies such as those performed by \citet{2018ARep...62..764T} have shown a linear relationship between radio continuum and infrared fluxes \rev{at 8 and 24~$\mu$m. \citet{2014ApJS..212....1A} quantified this correlation between 21~cm and 24~$\mu$m fluxes for a sample of 301 Galactic\hii~regions, reporting a power-law fit of $F_{24} = 0.03 \, F_{21\mathrm{cm}}^{0.90}$. While m}id-infrared emission in particular traces hot dust and photon-dominated regions via their PAH emission, \rev{and} radio emission traces ionised gas, a strong correlation suggests that both are good indicators of the presence of recently formed massive stars.

\begin{figure}
    \centering
    \includegraphics[width=\columnwidth]{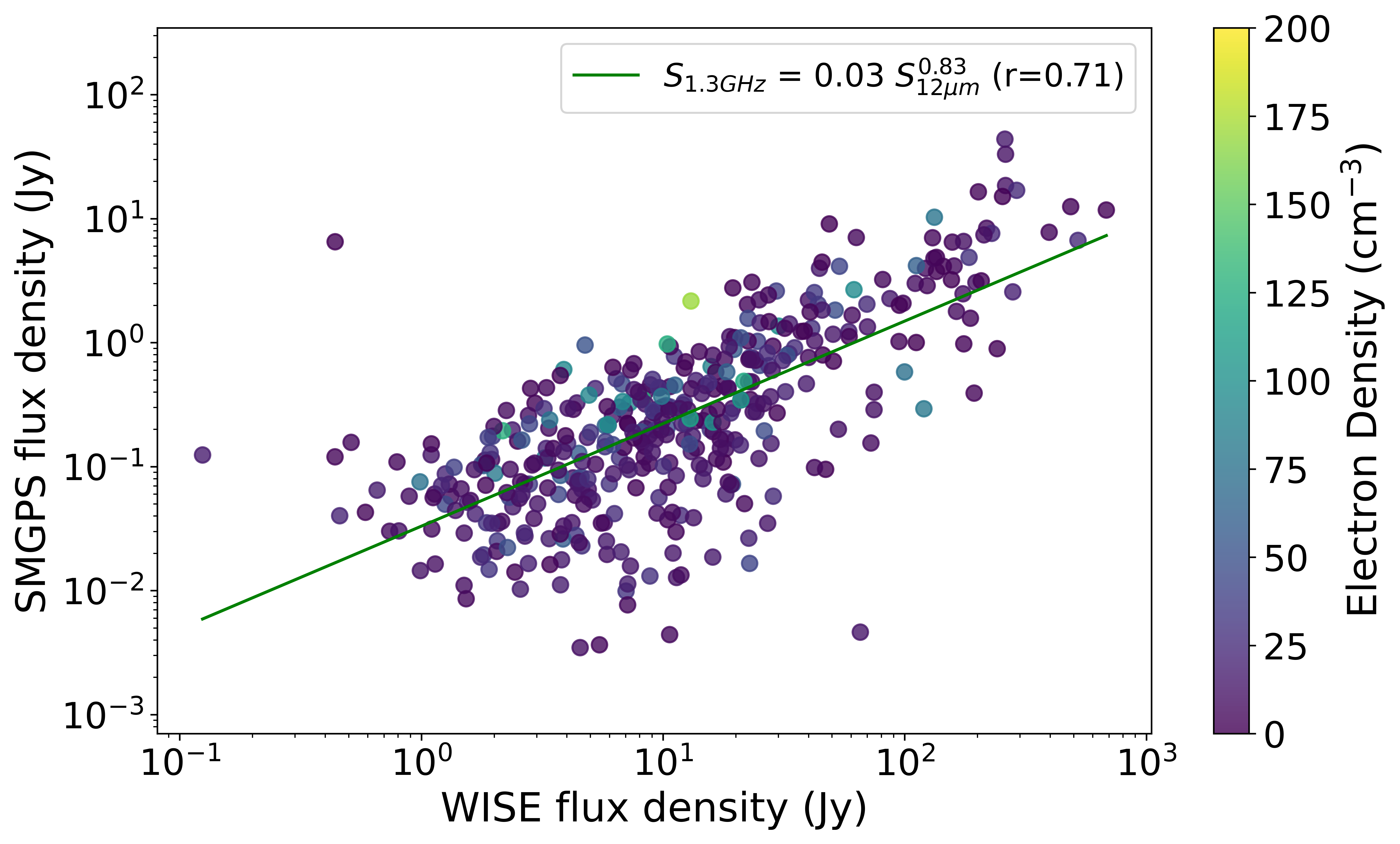}
    \caption{Correlation plot comparing the flux density of \hii~regions at 12 \(\mu \)m (WISE) and 1.3 GHz (SMGPS). This establishes the fundamental relationship between infrared dust emission and radio-ionised gas emission, validating their use as star-formation tracers. The line shows the best-fit linear trend \rev{with a gradient of $0.83\pm0.41$ (r=0.71)}. The electron density ($\text{cm}^{-3}$) of each ionised region, derived using equation \ref{eq:ne}, is represented by its colour coding, highlighting variations in gas concentration.}
    \label{fig:smgps_wise_flux}
\end{figure}

\rev{While previous studies, such as the one by \citet{2014ApJS..212....1A}, established these correlations using a sample of 301 \hii~regions, the high-sensitivity SMGPS allows us to test these relationships across a significantly larger sample of 1327 regions, offering more robust statistical constraints on the Galactic star-forming population.} We see a very similar relationship between the WISE mid-IR fluxes and the SMGPS. Figure~\ref{fig:smgps_wise_flux} shows the relationship between the flux densities at $12~\mu\mathrm{m}$ WISE and $1.3~\mathrm{GHz}$ SMGPS. 
\rev{We have chosen to compare our radio continuum against the 12~$\mu$m emission, firstly because the 12~$\mu$m band provides superior angular resolution ($\sim$6.5$''$) compared to the 22~$\mu$m band ($\sim$12$''$), allowing for a more precise morphological comparison with the high-resolution SMGPS data. Secondly, the 12~$\mu$m filter captures the strong PAH features that characterise the PDR, providing a clear physical boundary for the \hii~region's interaction with its natal molecular cloud.}
The electron density of each ionised region is mapped through the colour scale.
The yellow points represent regions at the upper \rev{end} of the classical \hii~electron density range \rev{($n_{\text{e}} > 200 \text{ cm}^{-3}$). While our survey is primarily sensitive to more evolved stages, when viewed in conjunction with literature samples of compact and ultracompact regions, it establishes a continuous evolutionary sequence across three orders of magnitude in physical size. This underscores that our sample effectively populates the previously undersampled transition between compact and classical diffuse environments, confirming a consistent scaling relationship across the known distribution of ionised environments.}
The best-fit linear trend illustrates a strong correlation between infrared dust emission and radio free–free emission. The scatter at the low-flux end of the distribution is likely dominated by low-signal-to-noise measurements rather than by intrinsic physical differences between sources \citep{2014ApJS..212....1A}. We also compared the $12\ \mu\text{m}$ infrared flux densities of WISE with the radio flux densities of SMGPS by calculating the ratios defined as $\log_{10}( F_{12\ \mu\text{m}} / F_{\text{radio}})$. Our analysis yields a \rev{mean} ratio of \rev{1.66 with a standard error of $\pm0.03$}.
\rev{To further evaluate this result, we categorised our sample according to the physical radius ranges defined by \citet{2017ApJ...846...64M}. We found mean ratios of $1.71 \pm 0.03$ for $1 < r < 5$~pc, $1.50 \pm 0.05$ for $5 < r < 10$~pc, and $1.28 \pm 0.08$ for $r > 10$~pc. These values are consistent with the results of \citet{2017ApJ...846...64M}, who reported ratios of 1.55, 1.67, and 1.54 for the same respective size ranges.}

Moreover, the similar angular resolution between WISE and SMGPS (6.5\arcsec\ at 12 $\mu$m vs 8\arcsec\ at 1.3 GHz) allows us to examine the spatial relationship between radio and infrared emission.
\rev{Figure \ref{fig:radii} illustrates the relationship between the angular radii of the \hii~regions as measured in the mid-IR (WISE) and radio (SMGPS) regimes. We find a significant positive correlation with a Pearson coefficient of $r = 0.91$ ($p < 0.001$). To address potential distance bias, where the correlation might be artificially inflated by the shared dependence of physical parameters on heliocentric distance, we calculated the partial Spearman rank correlation coefficient, controlling for distance. The resulting coefficient remains robust at $\rho_{p} = 0.84$ ($p < 0.001$). This confirms that the strong physical coupling between the PAH-emitting regions (mid-IR) and the ionised gas (radio) is intrinsic and not merely a byproduct of the distance to the sources.} This is remarkable given \emph{i)} the different measurement methods between WISE and SMGPS (by-eye vs algorithmic, respectively); and \emph{ii)} the different emission mechanisms involved \citep[PAH and hot dust for the mid-IR and free-free for radio][]{2014ApJS..212....1A}.

\rev{The close agreement between the mid-IR and radio radii is consistent with the established structure of \hii~regions, where the ionised volume is typically surrounded by a thin shell of dust and PAHs. While the radio emission traces the free-free radiation from the ionised gas within the Str\"omgren sphere, the mid-IR emission traces the interface of the PDR immediately outside the ionisation front. Our results show that across five orders of magnitude in size, the PDR remains closely coupled to the ionised volume. This confirms that even in more evolved, classical \hii~regions, the mid-IR morphology serves as a reliable proxy for the extent of the ionised gas, providing a unified view of the feedback scales from massive stars.}
Figure~\ref{fig:radii} provides a direct comparison of how these tracers outline the structure of Galactic $\textrm{H}\scriptstyle\mathrm{II}$ regions. The strong correlation observed suggests that for this population, both radio and infrared measurements provide a consistent characterisation of spatial extent. Notably, we do not observe a significant systematic effect where one tracer consistently overestimates the other. This close correspondence contrasts with studies of more compact $\textrm{H}\scriptstyle\mathrm{II}$ regions, such as those in the CORNISH survey \citep{Hoare2012_CORNISH_Design}, where radio extents are typically smaller than their mid-infrared counterparts. The agreement found here is likely attributable to the high surface brightness sensitivity of the SMGPS, which allows the detection of low-electron-density emission in the outer layers of classical $\textrm{H}\scriptstyle\mathrm{II}$ regions that might otherwise remain undetected in shallower radio observations.

\rev{For the 18 instances where a single WISE-detected \hii~region was resolved into multiple radio components in the SMGPS catalogue (comprising 37 distinct radio sources), we have excluded these from the subsequent correlation and size-comparison analyses. This ensures that the reported statistics are based on a pure sample of discrete, one-to-one morphological matches. By removing these blended sources, where a single larger infrared structure corresponds to multiple compact radio components, we also minimise artificial scatter in the $r \lesssim 1$~arcmin regime and confirm that the radio data allow for a definitive separation of individual star-forming units compared to the mid-infrared.}

\begin{figure}
    \centering
    \includegraphics[width=\columnwidth]{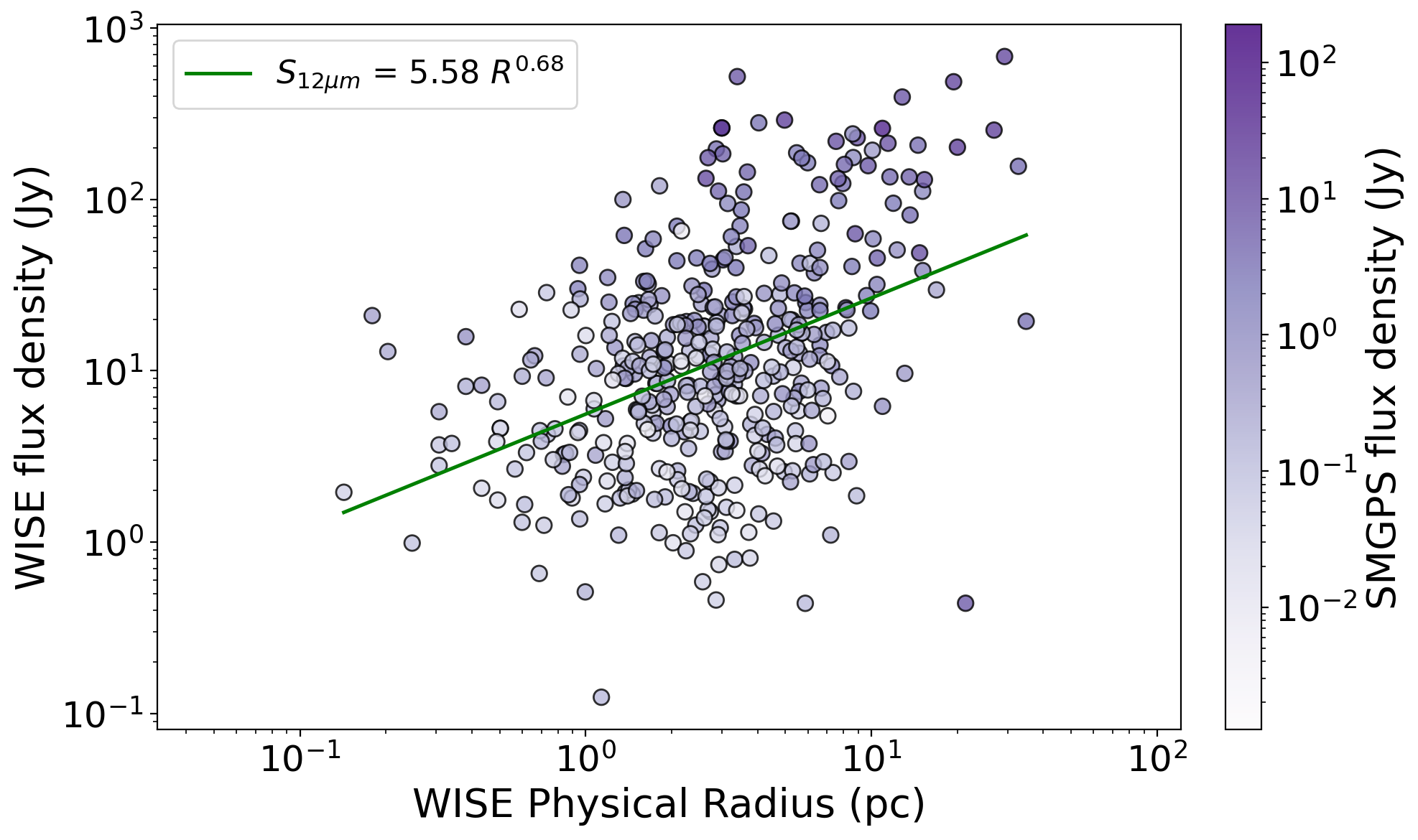}
    \caption{WISE 12 $\micro$m flux density as a function of physical size, colour-coded by the SMGPS flux density. The solid line shows the best-fit power-law relation in log-log space \rev{with a gradient of $0.68 \pm 0.07$}. A Spearman rank correlation test reveals \rev{weak} significant correlation (r = 0.38, p \rev{$< 0.001$}) between source size and flux, indicating that infrared luminosity is independent of physical extent for these \hii~regions.}
    \label{fig:wise_size_flux}
\end{figure}

Finally, we investigate whether there is a correlation between the WISE infrared flux densities and the physical radii of the $\text{H}\scriptstyle\mathrm{II}$ regions (see Figure~\ref{fig:wise_size_flux}). Our analysis reveals no significant dependence of the infrared flux on the physical size of the regions. This lack of correlation suggests that the mid-infrared luminosity is primarily driven by the ionising output of the central stellar source rather than the total volume of the expanded nebulae. This finding is in excellent agreement with \citet{2018ARep...62..764T}, who similarly observed that infrared flux levels are independent of the spatial extent of the region. This result underscores that the total luminosity traces the initial stellar energetics, while the physical size is a product of the evolution of the $\text{H}\scriptstyle\mathrm{II}$ region and the local ISM density.

\subsection{Galactic Distribution and Trends}
\label{galtrends}

\begin{figure*}
    \centering
    \includegraphics[width=\textwidth]{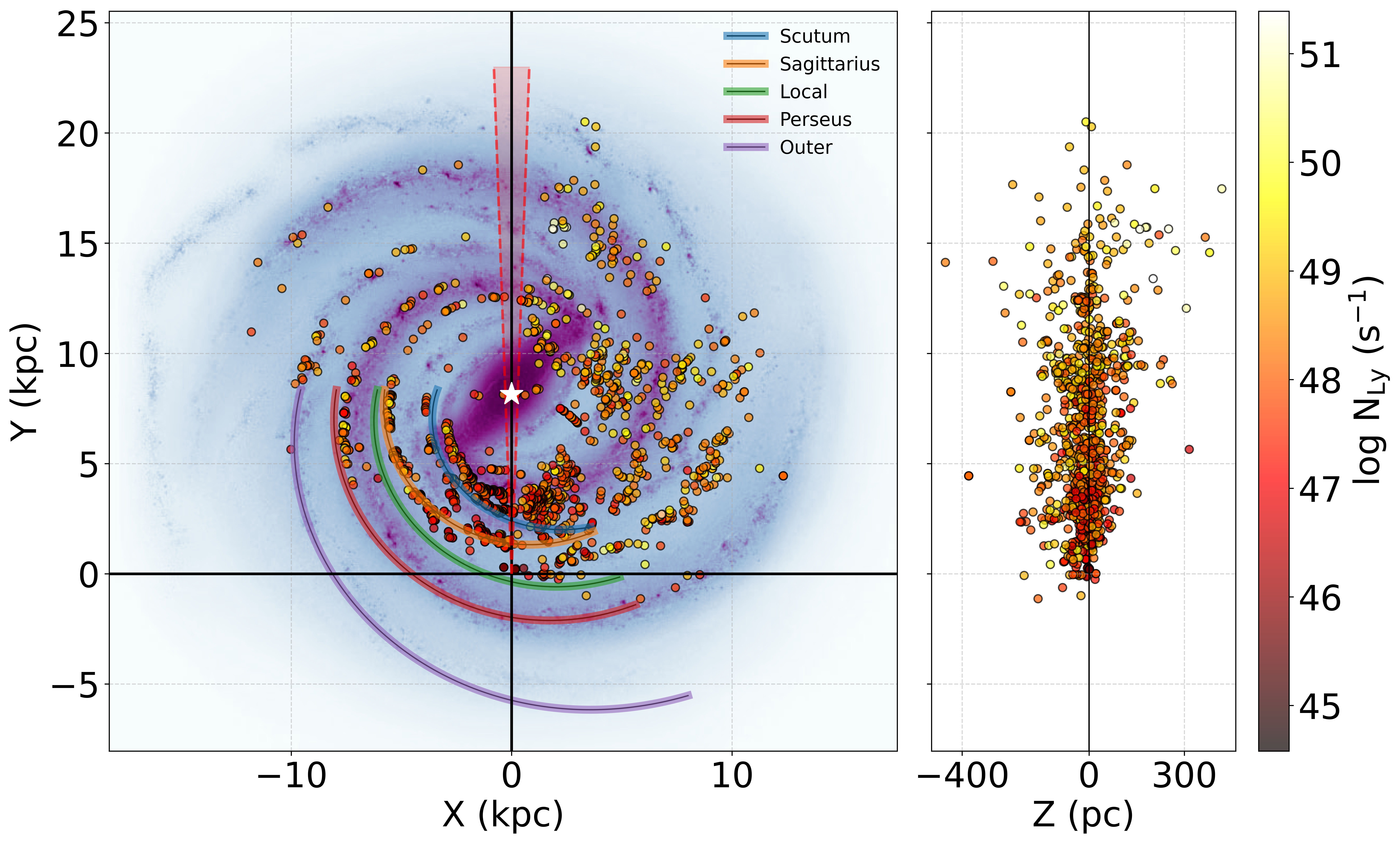}
    \caption{Left panel: The distribution of SMGPS {$\textrm{H}\scriptstyle\mathrm{II}$} regions within the Galactic plane, presented in Cartesian coordinates (X, Y) with the Sun at (0, 0) and the star indicates the location of the Sgr A* at the Galactic Centre. Distances to the {$\textrm{H}\scriptstyle\mathrm{II}$} regions are derived from the \citet{2019ApJ...885..131R} Bayesian calculator. The colour gradient represents the observed $\log(N_{\mathrm{Ly}})$, which is proportional to the Lyman continuum photon rate and indicative of the {$\textrm{H}\scriptstyle\mathrm{II}$} region's ionising flux. The distribution of the observed data is overlaid on an artistic representation of the Milky Way, which includes the spiral-arm model to show the approximate locations of the major spiral arms (see the legend for names). The shaded area bounded by dotted lines indicates the line-of-sight cone toward the Galactic Centre, which was excluded from this analysis. Right panel: The heliocentric radius distribution of {$\textrm{H}\scriptstyle\mathrm{II}$} region as a function of the Z axis, which is perpendicular to the Galactic plane. The plotted range spans from $-400$ to $300~\text{pc}$. Note that the $\text{Z}$ coordinates are not corrected for the Sun's displacement relative to the Galactic midplane.}
    \label{fig:hii_map}
\end{figure*}

By obtaining more accurate measurements of physical radii and heliocentric distances, we can better situate these regions within the Galactic context, effectively %mapping the population 
detecting individual \rev{\hii~regions}
even up to heliocentric distances of 21~kpc. \rev{The Galactic distribution of \hii~regions, shown in Figure \ref{fig:hii_map}, appears broadly consistent with the \citet{2019ApJ...885..131R} spiral arm model. We emphasise that this alignment is expected as it results from a circular argument: the Bayesian distance-estimation technique utilises the spiral arm model as a prior to resolve distance ambiguities. Therefore, Figure \ref{fig:hii_map}, should be interpreted not as an independent confirmation of the Milky Way’s structure, but rather as a demonstration of the internal consistency of our distance determinations within the \citet{2019ApJ...885..131R} framework. Despite this dependence, the results highlight the capability of radio observations to trace the large-scale structures of the Milky Way, such as the Carina–Sagittarius arm, which is a prominent structure in the inner Galaxy where massive star formation typically occurs. The ability to trace spiral arms with {$\textrm{H}\scriptstyle\mathrm{II}$} regions is well-established, as {$\textrm{H}\scriptstyle\mathrm{II}$} regions are signposts of recent and active massive star formation, which typically occurs along the dense molecular gas lanes of spiral arms \citep[e.g.][]{1976A&A....49...57G}. The observed deviations from the expected rotation curve likely stem from non-circular motions within spiral arms or inherent uncertainties in kinematic distance modelling. According to \citet{1994ApJS...91..659K}, the combination of peculiar motions and distance ambiguities between near and far measurements in the inner Galaxy obscured the underlying spiral arm structure, making it difficult to trace these features using solely {$\textrm{H}\scriptstyle\mathrm{II}$} regions. The vertical distribution of the sample (Figure \ref{fig:hii_map}, right panel) shows that the majority of regions are concentrated within $\pm$100~pc of the Galactic midplane. This tight confinement to the disk is characteristic of young, high-mass star-forming regions and indicates that our sample successfully traces the thin-disk component of the Galaxy, where the densest molecular gas resides.}

\begin{figure}
    \centering
    \includegraphics[width=\columnwidth]{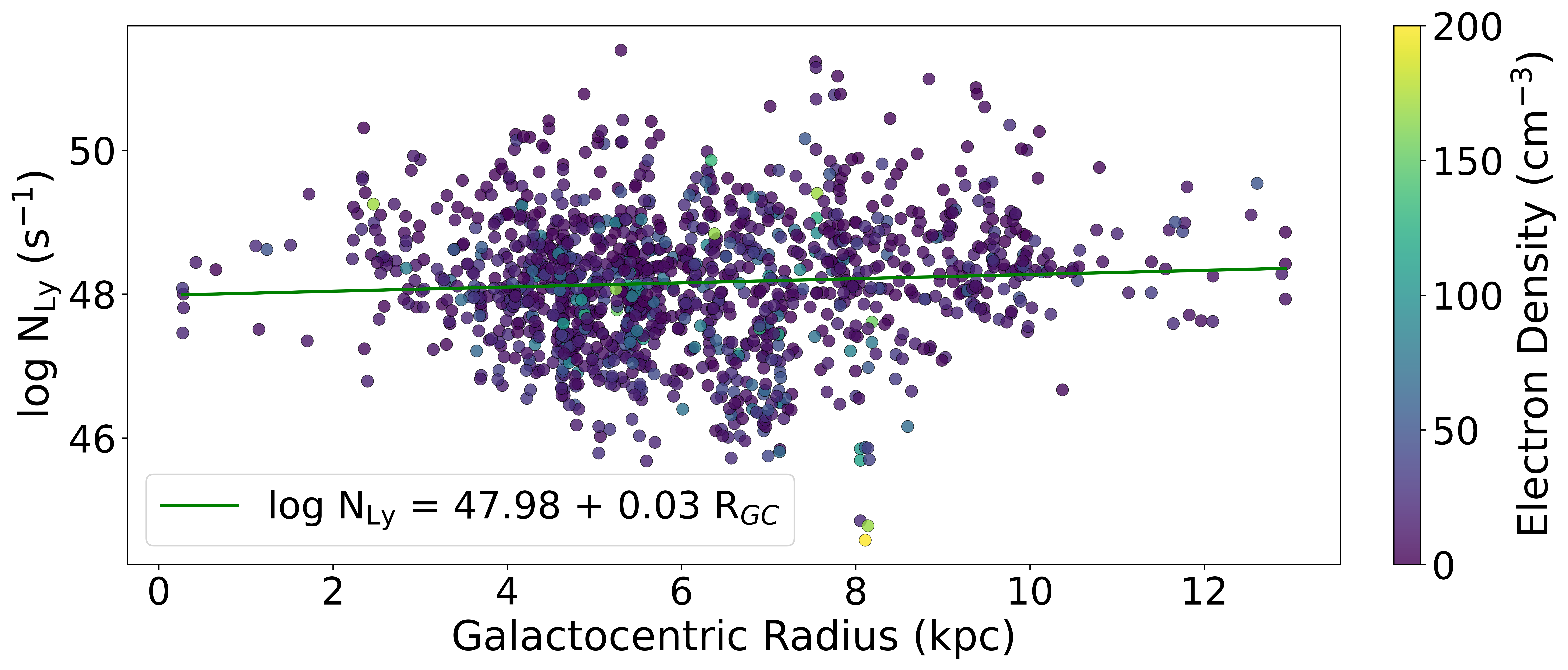}
    \caption{Ionising photon flux ($\log \textrm{N}_{Ly}$ [s$^{-1}$]) for SMGPS {$\textrm{H}\scriptstyle\mathrm{II}$} as a function of galactocentric distance. Points are colour-coded by electron density. A Spearman rank correlation test reveals no significant correlation (r = 0.06, p = 0.04) between log N$_{Ly}$ and distance from the Galactic centre\rev{, with a very weak positive gradient of $0.03 \pm 0.01$}. The colour-coding indicates the electron density (cm$^{-3}$).}
    \label{fig:nly_gc}
\end{figure}

Moreover, no significant correlation in $N_{\mathrm{Ly}}$ is observed with the Galactocentric distance across the sample, as shown in Figure \ref{fig:nly_gc}, suggesting that the observed trend is primarily governed by local physical conditions rather than by the large-scale Galactic environment.
The absence of a significant correlation between Lyman continuum flux and Galactocentric distance found \rev{in this study} is consistent with previous \rev{findings}. In particular, \rev{\citet{2019MNRAS.487.1057D}} reported a very weak negative gradient of \rev{$-0.06 \pm 0.02$ kpc$^{-1}$} in the Lyman-photon flux with Galactocentric radius. \rev{This slope is consistent with zero, and the relationship was characterised by a non-significant correlation ($p > 0.05$) and substantial scatter. While previous literature sometimes attributed this scatter to the assumption of optically thin emission, it is unlikely that our SMGPS sample contains significant populations of optically thick \hii~regions given their physical sizes. Instead, the observed scatter and potential underestimation of $N_{\rm Ly}$ are more likely driven by the absorption of ionising photons by dust within the \hii~regions and the leakage of ionising radiation into the surrounding interstellar medium \citep{2001ApJ...549..979K, 2012AJ....144..173D}.}

\begin{figure}
    \centering
    \includegraphics[width=\columnwidth]{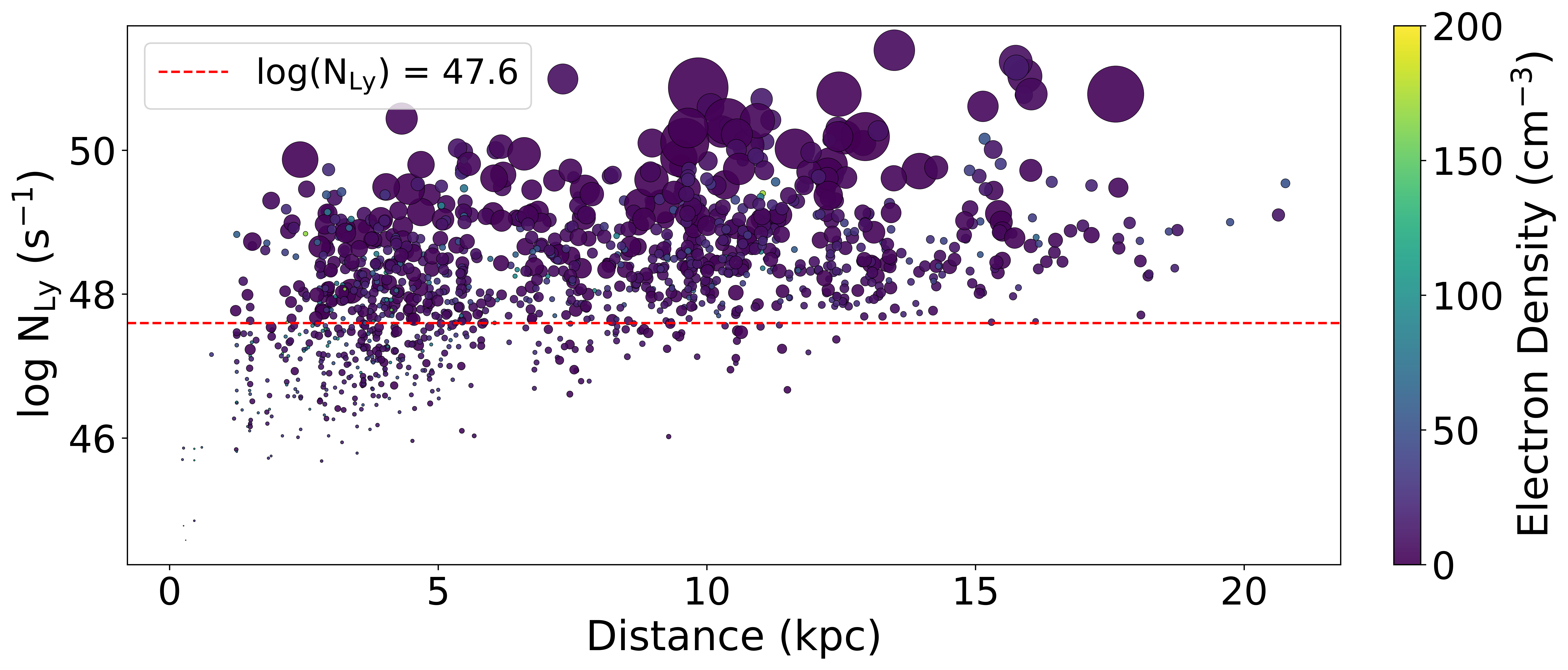}
    \caption{Ionising photon flux ($\log N_{\mathrm{Ly}}$ [s$^{-1}$]) as a function of heliocentric distance for {$\textrm{H}\scriptstyle\mathrm{II}$} regions in the Galactic plane. The dashed horizontal line at log($N_{\mathrm{Ly}}$) = 46.8 marks the RRL completeness detection limit, beyond which most sources are detected due to their high Lyman-photon rate. Marker sizes are proportional to the physical sizes of the regions, and colours indicate electron density (cm$^{-3}$).}
    \label{fig:nly_distance}
\end{figure}

Figure \ref{fig:nly_distance} shows the distribution of Lyman fluxes and the physical sizes of our \hii\ region sample with heliocentric distance. The N$_{\rm Ly}$ distribution shows a clear Mal\rev{m}quist bias with distance, where the lower $N_{\rm Ly}$ \hii\ regions are only detected at nearer distances. We propose a nominal completeness limit of log $N_{\mathrm{Ly}} \approx 47.6~ \textrm{s}^{-1}$ for our sample up to a distance of \rev{$\sim$15 kpc (Figure \ref{fig:nly_distance})}. It is important to note that this completeness limit is likely driven by the sensitivity of the  RRL observations rather than the continuum sensitivity of the SMGPS (see Section \ref{sec:properties}).  
This means that many lower-luminosity {$\textrm{H}\scriptstyle\mathrm{II}$} regions, i.e. those ionised by less massive B-type stars or more evolved and diffuse regions farther away, are likely to be missed.
It is also important to consider that some of these “missing sources” at lower photon fluxes may in fact be associated with {$\textrm{H}\scriptstyle\mathrm{II}$} regions previously classified as radio quiet. In \citet{2025A&A...695A.144B}, radio emission was detected towards 759 such “radio-quiet” {$\textrm{H}\scriptstyle\mathrm{II}$} regions, demonstrating that a non-negligible fraction of these objects host faint or compact radio emission below earlier detection thresholds. Additional sources may also be present in the SMGPS compact source catalogue \citep{2026MNRAS.546f1849M} or among {$\textrm{H}\scriptstyle\mathrm{II}$} region candidates in \citet{2014ApJS..212....1A} that lack RRL detections. This will be explored further in a forthcoming SMGPS study of {$\textrm{H}\scriptstyle\mathrm{II}$} region candidates (Chibueze et al., in prep.).

Ultra-compact (UC) and compact {$\textrm{H}\scriptstyle\mathrm{II}$} regions represent earlier evolutionary stages and often have a higher surface brightness despite their smaller physical sizes \citep{2013MNRAS.435..400U}. If these compact sources were below the resolution limit of WISE, but still detected by radio observations, they might contribute to the smaller angular-radii population seen by SMGPS. However, if their photon fluxes are below the 47.6 $\textrm{s}^{-1}$ threshold, they would still be challenging to characterise \rev{using RRLs alone. While velocity information for C\hii~regions can often be recovered from molecular line observations of their natal environments \citep[e.g.,][]{2013MNRAS.435..400U}, more evolved and extended regions like those in our sample often lack such clear molecular associations, making the RRL detection limit a more critical constraint.} The bias of RRL observations toward detecting nearby or high-luminosity sources is a direct consequence of the fact that RRL brightness scales proportionally with the radio continuum brightness. For a given electron temperature, the line-to-continuum ratio remains relatively constant; consequently, lower-luminosity B-type {$\textrm{H}\scriptstyle\mathrm{II}$} regions, which exhibit faint continuum emission, produce comparably weak RRL signals \citep{2002ASSL..282.....G}. When these sources are located at greater distances, their already faint emission often falls below the survey's sensitivity threshold, preventing reliable velocity measurements. This incompleteness at the faint end of the luminosity function ultimately limits our ability to derive kinematic distances and accurate ionising photon fluxes for the most distant, lower-excitation regions.
Moreover, detected RRL sources exhibit a clear tendency to cluster, particularly among neighbouring regions. This apparent clustering is likely the result of the detection bias, whereby at greater distances similar groups can blend together or fall below the detection threshold, while nearby regions often contain multiple \hii~regions that can be individually detected and resolved. The recovery of the entire population of ionising sources, improved completeness at lower flux densities, and improved constraint of the initial mass function and early evolution of massive stars will all depend on future studies that combine SMGPS with complementary infrared, millimetre, and higher-frequency radio surveys, particularly those that can detect fainter RRLs or compact thermal radio emission.

\subsection{\rev{Case Studies: Probing Evolutionary Diversity}}

The sample presented herein highlights the morphological diversity encountered within the SMGPS dataset and underscores the survey's ability to dissect complex star-forming environments. The case of {\tt{G003.265-00.053}}, illustrated in Figure \ref{fig:smultiplot}, exemplifies the ability of MeerKAT to trace the large-scale ($\approx45.8$ pc) continuum emission associated with 'classical' diffuse {$\textrm{H}\scriptstyle\mathrm{II}$} regions. The identification of spatially distinct groups of candidate ionising stars within this extended structure suggests ongoing high-mass star formation across the complex. The detection of embedded sub-regions like {\tt{G003.349-00.079}}, whose infrared emission aligns with the larger structure, confirms the physical association and hierarchical nature common in complexes of the {$\textrm{H}\scriptstyle\mathrm{II}$} region \citep{2002ARA&A..40...27C}. The sensitivity of MeerKAT is important in capturing the full extent of such extended emission, providing constraints on the radiated ionising photons and the interaction of {$\textrm{H}\scriptstyle\mathrm{II}$} regions with the surrounding interstellar medium. In contrast, {\tt{G003.439-00.349}} represents a compact {$\textrm{H}\scriptstyle\mathrm{II}$} region ($\approx0.5$ pc), consistent with ionisation by a single B0-type young massive star. Such objects represent an early phase in massive star evolution, where the ionised gas remains confined mainly by the dense natal molecular cloud \citep{1989ApJS...69..831W}. The resolution of SMGPS, while not completely resolving the internal structure here, confirms its compact nature and differentiates it from more evolved and extended regions. These contrasting examples underscore the breadth of evolutionary stages probed by SMGPS, from deeply embedded compact sources to fully evolved extended complexes, asserting the critical role of the survey in constructing a comprehensive view of star formation across the Galactic plane.

\section{Summary and Conclusion}
\label{conclusion}

We leveraged the high sensitivity of the MeerKAT telescope to detect faint and extended Galactic ionised structures overlooked by previous targeted radio surveys. This capability is valuable for studying the large-scale morphology, uncovering the full extent of {$\textrm{H}\scriptstyle\mathrm{II}$} regions, especially classical {$\textrm{H}\scriptstyle\mathrm{II}$} regions with significant diffuse emission. We computed the 1.3 GHz radio continuum Lyman-photon flux and electron densities toward 1,327 {$\textrm{H}\scriptstyle\mathrm{II}$} regions, significantly expanding the number of WISE \hii\ regions with high angular resolution radio continuum observations. We determined the spectral types of the candidate ionising stars of these regions using the distance estimates from the Bayesian distance calculator of \citet{2019ApJ...885..131R}, which incorporates the Galactic structure as a priori to resolve kinematic ambiguities. The derived stellar spectral types range from B2 to O4, with a peak around O7.5, equivalent to log Lyman-photon flux of 48.18 s$^{-1}$.

\begin{itemize} %\itemsep{8 pts}

\item The Lyman-photon flux density distribution indicates that a substantial number of low-luminosity objects are missing from our sample, which do not have RRL detections. 
%therefore suggesting a $100\%$ detection cut-off at log $N_{\mathrm{Ly}}$  $\sim46.8~ \textrm{s}^{-1}$. 
Therefore, high-sensitivity radio\rev{-}spectroscopic observations focusing on RRLs and molecular tracers are needed to characterise the kinematics of the lower\rev{-}luminosity sample.

\item We confirm that the constant evolution of N$_{\rm Ly}$ with physical radius \citep[as observed by][for hypercompact to compact \hii\ regions]{2025MNRAS.538.2267P} extends into the classical regime.

\rev{\item We find a robust correlation between radio and mid-infrared angular radii (slope $m = 1.15 \pm 0.02$). The slight excess in radio extent suggests that in more evolved H$\scriptstyle\mathrm{II}$ regions, ionised gas (1.3 GHz) expands beyond the 12 $\mu$m dust emission.}

\item The lack of correlation between mid-infrared flux and physical radius confirms that while total luminosity traces the ionising output of the central star, the physical extent is governed by independent dynamical expansion and the local interstellar environment.

\item Our findings demonstrate that SMGPS provides a comprehensive view of evolved \hii~regions by capturing the large-scale diffuse emission, thereby providing a more representative census of the population of classical {$\textrm{H}\scriptstyle\mathrm{II}$} regions.

\end{itemize}

The future is ripe for further exploitation of the SMGPS dataset for \rev{\hii~region} studies. Efforts are underway to recalibrate and re-image SMGPS (Ramaila et al., in prep.), which will enable the accurate calculation of in-band spectral indices, polarimetry and greater sensitivity to faint extended emission. Currently, in-band spectral index studies are mainly limited by the SMGPS imaging strategies \citep{2024MNRAS.531..649G}. Furthermore, these results establish a foundation for forthcoming large‐scale surveys of the Galactic Plane using MeerKAT and, ultimately, the Square Kilometre Array (SKA). In particular, planned observations at higher frequencies $\sim 3$ GHz and SKA Band 5b will enable more sensitive and spatially resolved studies of Galactic ionised gas structures. The $\text{SKA}$ Band $5\text{b}$ ($\SI{10}{\giga\hertz}-\SI{15}{\giga\hertz}$) is ideally suited for spectral line analysis, including $\text{RRL}$ detection, which will provide new constraints on the electron temperature and precise gas dynamics (Traficante et al., in prep.).

\section*{Acknowledgements}

The MeerKAT telescope is operated by the South African Radio Astronomy Observatory, a facility of the National Research Foundation, an organisation under the Department of Science and Innovation, which runs the MeerKAT telescope. We appreciate the efforts of the entire MeerKAT team at SARAO to build and operate the MeerKAT. OMS's research is supported by the South African Research Chairs Initiative of the Department of Science and Technology and National Research Foundation (grant No. 81737). MAT acknowledges support from STFC grant awards ST/R000905/1 and ST/W00125X/1. The Centre for Radio Astronomy Techniques and Technologies at Rhodes University provided access to their Computing facilities for data processing and storage.

\section*{Data availability}
All SMGPS DR1 data products are available through the DOI: \url{https://doi.org/10.48479/3wfd-e270}. The raw visibilities are available on the SARAO Data Archive (\url{https://archive.sarao.ac.za}) under the project code SSV-20180721-FC-01. The SMGPS Extended Source Catalogue \citep{2025A&A...695A.144B} is available at \url{https://doi.org/10.48479/t1ya-na33}. In addition, a supplementary interactive web resource providing visualisations of the SMGPS {\(\textrm{H}\scriptstyle \mathrm{II}\)} regions, including image cutouts and the full source catalogue, is available at
\url{https://athanaseus.github.io/hii}.

\bibliographystyle{mnras}
\bibliography{references} % if your bibtex file is called example.bib
%\end{thebibliography}

%%%%%%%%%%%%%%%%%%%%%%%%%%%%%%%%%%%%%%%%%%%%%%%%%%
%%%%%%%%%%%%%%%%% APPENDICES %%%%%%%%%%%%%%%%%%%%%
\appendix

\section{Structure of the Source Catalogue}
\label{appendix}

\begin{table*}
\centering
\caption{Format of the SMGPS extended $\textrm{H}\scriptstyle\mathrm{II}$ regions catalogue (Appendix Table A1)}
\label{tab:source_properties}
\begin{threeparttable}
\begin{tabular}{cccl}
\hline
Col. Num. & Name & Unit & Description \\
\hline
1 & SMGPS Name & -- & Source identifier (GLLL.lll$\pm$BB.bbb) as identified by SMGPS. \\
2 & WISE Name & -- & Name identifier as in the WISE catalogue of $\textrm{H}\scriptstyle\mathrm{II}$ regions. \\
3 & $F_{12\mu\text{m}}$ & Jy & WISE 12~$\mu$m flux density. \\
4 & $v_{\text{LSR}}$ & km/s & Velocity with respect to the Local Standard of Rest. \\
5 & $D \pm \delta D$ & kpc & Estimated distance and associated error using \citet{2019ApJ...885..131R} calculator. \\
6 & $R_{\text{WISE}}$ & pc & WISE physical radius in parsecs. \\
7 & $F_{\text{radio}} \pm \delta F$ & Jy & SMGPS flux density and error measured at 1.3~GHz. \\
8 & $R_{\text{radio}}$ & pc & SMGPS physical radius in parsecs. \\
9 & $n_{\text{e}}$ & cm$^{-3}$ & Derived electron density of the ionised region. \\ 
10 & $\log N_{\text{Ly}}$ & s$^{-1}$ & Logarithmic Lyman continuum photon flux. \\
11 & Spec. Type & -- & Stellar classification based on \citet{1973AJ.....78..929P} calibrations. \\
\hline
\end{tabular}
\begin{tablenotes}
\small
\centering
\item \textbf{Note:} The complete machine-readable version of this catalogue is available in the online journal. This table provides a description of the parameters included in the final data product.
\end{tablenotes}
\end{threeparttable}
\end{table*}

% Don't change these lines
\bsp	% typesetting comment
\label{lastpage}
\end{document}